\documentclass[useAMS,usenatbib]{mn2e}

\usepackage{graphicx}
\usepackage{dblfloatfix}

\bibpunct{(}{)}{;}{a}{}{,} 

\newcommand\kms{{\rm \,km\,s^{-1}}}
\newcommand\kpc{{\rm \,kpc}}

\renewcommand\d {{\rm d}}
\newcommand\yr{{\rm \,yr}}

\newcommand {\sm}{\rm\,M$_\odot$}

\def\FVFPS{{\sc fvfps\,}}
\newcommand{\vx}{v_x}
\newcommand{\vy}{v_y}
\newcommand{\vz}{v_z}
\newcommand{\vr}{v_{\rm r}}
\newcommand{\vt}{v_{\rm t}}
\newcommand{\RS}{R_{\rm S}}

\newcommand{\rcDM}{r_{\rm c,DM}}
\newcommand{\rtDM}{r_{\rm t,DM}}
\newcommand{\rcstar}{r_{\rm c,*}}
\newcommand{\rhostar}{\rho_*}
\newcommand{\rhoDM}{\rho_{\rm DM}}
\newcommand{\Izero}{I_0}
\newcommand{\Mstar}{M_*}
\newcommand{\Mtot}{M_{\rm tot}}
\newcommand{\MDM}{M_{\rm DM}}
\newcommand{\Msun}{M_{\odot}}

\newcommand{\apjl}{ApJL }

\title[The effects of tides on Fornax]{The effect of tides on the Fornax dwarf spheroidal galaxy}
\author[Battaglia, Sollima, Nipoti]{Giuseppina Battaglia$^{1,2}$\thanks{E-mail:
gbattaglia@iac.es}, Antonio Sollima$^{3}$ and 
Carlo Nipoti$^{4}$\\
$^{1}$Instituto de Astrofisica de Canarias, calle Via Lactea s/n, 38205 La Laguna, Tenerife, Spain\\
$^{2}$Universidad de La Laguna, Dpto. Astrofisica, 38206 La Laguna, Tenerife, Spain\\
$^{3}$INAF - Osservatorio Astronomico di Bologna, via Ranzani 1, 40127 Bologna, Italy\\
$^{4}$Department of Physics and Astronomy, University of 
Bologna, Viale Berti Pichat 6/2, 40127 Bologna, Italy}
\begin{document}

\date{Accepted 2015 September 07. Received 2015 September 03; in original form 2015 August 06.}

\pagerange{\pageref{firstpage}--\pageref{lastpage}} \pubyear{2015}

\maketitle

\label{firstpage}

\begin{abstract}
Estimates of the mass distribution and dark-matter (DM) content of
dwarf spheroidal galaxies (dSphs) are usually derived under the
assumption that the effect of the tidal field of the host galaxy is
negligible over the radial extent probed by kinematic data-sets.  We
assess the implications of this assumption in the specific case of the
Fornax dSph by means of $N$-body simulations of a satellite orbiting
around the Milky Way.  We consider observationally-motivated orbits
and we tailor the initial distributions of the satellite's stars and
DM to match, at the end of the simulations, the observed structure and
kinematics of Fornax. In all our simulations the present-day
  observable properties of Fornax are not significantly influenced by
  tidal effects.  The DM component is altered by the interaction with
  the Galactic field (up to 20\% of the DM mass within 1.6 kpc is
lost), but the structure and kinematics of the stellar component are
only mildly affected even in the more eccentric orbit (more than 99\%
of the stellar particles remain bound to the dwarf).  In the
simulations that successfully reproduce Fornax's observables, the
dark-to-luminous mass ratio within 1.6 kpc is in the range $5-6$, and
up to $16-18$ if measured within 3 kpc.
\end{abstract}


\section{Introduction}
\label{sec:intro}

Over the years, the manyfold growth in the size of spectroscopic
samples of individual stars has allowed the determination of accurate
and spatially extended line-of-sight (LOS) velocity-dispersion
profiles for each of the Milky Way (MW) ``classical'' dwarf spheroidal
galaxies (dSphs\footnote{We refer to the dSphs known prior to the SDSS
  as ``classical'' dSphs, while those discovered afterwards as
  ``ultra-faint dwarfs''.}; e.g. Kleyna et al. 2001; Koch et al. 2007;
Walker et al. 2007; Battaglia et al. 2008; \citealt{Mun06}; Battaglia
et al. 2011; Walker 2013) as well as of higher moments of the LOS
velocity distribution (e.g. \citealt{Lok09}; Breddels et al. 2013;
Breddels \& Helmi 2013, hereafter BH13; \citealt{Ric14}).  Equilibrium
dynamical modeling of these data-sets leads to the finding that MW
dSphs exhibit some of the most extreme dynamical mass-to-light (M/L)
ratios known to-date, up to several 100s (M/L)$_{\odot}$. This has
triggered a large body of works attempting to assess the dark matter
(DM) content and distribution of these galaxies (for recent reviews
see Battaglia, Helmi \& Breddels 2013; Strigari 2013; Walker 2013),
which is an important piece of information to understand the nature of
DM particles (e.g. Ackermann et al. 2014) and to test the predictions
of cosmological models of galaxy formation, for instance regarding the
mass function of luminous satellites around MW-like systems
(e.g. Klypin et al. 1999; Moore et al. 1999) and the effects of
baryons on the DM distribution (e.g. \citealt{Nav96,Nip15}, Zolotov et
al. 2012).

However, it has always been a matter of debate whether the tidal
interaction experienced by these small galaxies during their orbital
evolution around the MW may have impacted their dynamical status to
the extent of significantly altering estimates of their current DM
content and distribution. Observational evidence of on-going tidal
disturbance in some M31 satellite galaxies is provided by the clear
presence of tidal tails or strong structural distortions
(e.g. \citealt{Cho02}; McConnachie et al. 2010; Crnojevic et
al. 2014).  Around the MW, beside the spectacular case of
Sagittarius (e.g. \citealt{Iba94}), several ``ultra-faint'' dwarf
galaxies exhibit distorted or very elongated morphologies (e.g. the
Hercules dSph; \citealt{Rod15}) however, among the ``classical''
dSphs, only Carina shows clear signs of tidal disturbance in its
structural properties (\citealt{Mun06}; Battaglia et al.  2012;
McMonigal et al. 2014). Given the difficulties in detecting
low-surface brightness tidal features from stellar counts in objects
as faint and extended on the sky as the MW dSphs, it is still not
excluded that such non-detections may be due to the lack of
appropriately deep and spatially extended photometric data-sets.

A tumultuous past, marked by strong mass loss due to interactions with
the host, is expected also from a theoretical point of view, in order
to explain the present-day properties of the observed population of MW
satellite galaxies in a well-motivated cosmological framework
(e.g. Sawala et al. 2014; Barber et al. 2015), and even more so in
those models that explain dSphs as the result of ``tidally stirred''
dwarf disc galaxies (e.g. Mayer et al. 2001, Kazantzidis et al. 2011,
\citealt{Tom15}).

In general, as long as the satellite galaxy retains a bound core, the
LOS velocity dispersion of the stars in the central regions remains a
good indicator of the maximum circular velocity and bound mass
(e.g. \citealt{Mun08}; \citealt{Pen08}; Klimentowski et
al. 2009; Kazantzidis et al. 2011). At larger projected radii,
however, stars originating in the tidal tails may contaminate
kinematic samples of individual stars in MW dSphs, inflating their
measured velocity dispersions and significantly altering the DM
content and distribution inferred from an equilibrium dynamical
analysis (e.g. Klimentowski et al. 2007, Read et al. 2006;
\citealt{Mun08}).

The amount of tidal disturbance experienced is obviously dependent on
the orbital history of the satellite. However, even if an object has
suffered a large degree of mass loss, this does not necessarily imply
lack of current dynamical equilibrium.  Tidally perturbed stars
progressively become unbound and are eventually dispersed, with the
object eventually settling in a new equilibrium configuration
\citep{Pen09}. Also, the degree of contamination of kinematic
data-sets from stars originating in tidal tails can vary along the
orbit, as a consequence of the varying orientation of the inner
regions of the tidal tails with respect to the LOS (e.g. Klimentowski
et al. 2009). Given the difficulty in unambiguously identifying
unbound stars lingering close to the main body of the MW dSphs, it is
quite possible that, depending on the specific orbital history and
current location along the orbit, the observed kinematics of stars is
a faithful tracer of their underlying potential in some of the MW
dSphs and not in others.

$N$-body simulations tailored to reproduce the present-day position
and systemic velocity, as well as the observed structural and internal
kinematic properties of {\it specific} MW dSphs appear particularly
well suited to explore this issue. Such an approach has been for
example applied to Carina (\citealt{Mun06}) and Leo~I (Sohn et
al. 2007, \citealt{Mat08}), for which it was shown that modeling
these galaxies as tidally disturbed systems initially embedded in
``mass-follows-light'' DM halos, rather than in extended halos, yields
a good match to the data. The inferred DM content is much lower than
what inferred from a Jeans analysis in the hypothesis of dynamical
equilibrium (from 24(M/L)$_{\odot,V}$ in Koch et al. 2007 to $\sim$5
(M/L)$_{\odot,V}$ in Sohn et al. 2007, for Leo~I; from 80
(M/L)$_{\odot,V}$ in Walker et al. 2007 down to $\sim$40
(M/L)$_{\odot,V}$ in \citealt{Mun08} for Carina). The orbits
that provide the best match to the data have small perigalacticons and
high eccentricity. While for Carina this is in good agreement with the
inferences from the measured proper motion \citep{Pia03}, it is
likely that Leo~I orbit has a larger perigalacticon distance,
91$\pm$36 kpc \citep{Soh13}. Hence the impact of tidal interaction
with the MW might have been significantly over-estimated for this
galaxy.

For investigations of the dynamical status of the MW dSphs, it is
therefore advisable to focus on modeling objects placed on
observationally motivated orbits. In this work, we take under
consideration the Fornax dSph. There is consensus that this object is
moving on a rather external orbit around the MW (Dinescu et al. 2004;
\citealt{Pia07}, hereafter P07; \citealt{Men11}), so Fornax is a
good candidate of dSph that might be close to equilibrium. Our aim is
to quantify the effects of tides on Fornax's structure and
kinematics, by means of $N$-body simulations following the evolution
in the MW of a satellite dwarf galaxy, similar in structure to Fornax,
on orbits that are consistent with the current observational
constraints, including Fornax's proper motions.  The Fornax dSph is
also one of the best studied MW dSphs, hence we can then rely on a
comprehensive set of observational properties against which to compare
the models.

We review the relevant set of observables in Section~\ref{sec:obs}. In
Section~\ref{sec:setup} we describe the set-up of the $N$-body
simulations, whose results are shown in Section~\ref{sec:results}. We
conclude with a summary in Section~\ref{sec:summary}.

\section[]{The Fornax dwarf spheroidal galaxy: observed properties}\label{sec:obs}

The $N$-body simulations used in this work have been set up in order
to reproduce the main observables of the Fornax dSph.  Concerning the
structural properties of the stellar component, we focus on
reproducing the measured stellar mass and observed surface number
density profile, while for the internal kinematics we focus on the
observed LOS velocity-dispersion profile.  The simulated object is
also required to have present-day projected position, heliocentric
distance, systemic radial velocity and systemic proper motion similar
to the observed ones.

The Fornax dSph appears to have a regular, well-behaved 2D stellar
distribution, with constant ellipticity $\epsilon\simeq 0.3$, as
revealed by spatially extended photometric data reaching down to the
base of the red giant branch (RGB; e.g. Battaglia et al. 2006; see
Bate et al. 2015 for a recent, panoramic view over a 25 deg$^2$ region
around Fornax based on VST/ATLAS data).  The only sign of irregularity
at large spatial scales was claimed by Coleman et al. (2005), who
detected a shell-like feature outside Fornax's nominal King tidal
radius ($\simeq 3.25$ kpc) in the distribution of RGB stars along the
minor axis. The deeper imaging survey by Bate et al. (2015) showed
that this feature is a mis-identified over-density of background
galaxies.  In the inner regions the stellar distribution shows
departures from elliptical symmetry \citep{Irw95}, which are very
likely due to the asymmetric spatial distribution of the young main
sequence stars (see \citealt{Ste98}; Battaglia et al. 2006; de Boer et
al. 2013).  There are also a number of small stellar substructures in
the spatial distribution and kinematics of the stars within 0.3 deg
(Coleman et al. 2004; Battaglia et al. 2006; \citealt{Ols06}; de Boer
et al. 2013), some of which may be related to the accretion of a
smaller stellar system and others simply to patchy recent star
formation. All of the above signs of irregularity do not require
  the occurrence of significant tidal effects.

Fornax's distance modulus has
been derived from various techniques - optical photometry of the
horizontal branch, the tip of the RGB, RR Lyrae stars, red clump (RC)
stars, as well as from near-infrared photometry of the RC and RGB tip
(\citealt{Pie09} and references therein). These are all in good
agreement within the typical standard and systematic errors, yielding
an average value of $\sim$140 kpc, very close to the value here
adopted, i.e. $d$= 138$\pm$8 kpc \citep{Sav00}.

We take as reference the ellipticity $\epsilon$, position angle (P.A.)
and observed surface number-density profile from \citet{Bat06}, who
used relatively deep ESO/WFI V \& I photometry extending out to
Fornax's nominal King tidal radius to study the structure of this
galaxy. The observed surface number density as a function of
circularized\footnote{In the literature the observed properties of
  Fornax are often reported as functions of the semi-major axis $a$
  (also called ``elliptical radius''). In this work, in both
  observations and simulations, we always refer to the circularized
  radius $R\equiv\sqrt{ab}=a\sqrt{1-\epsilon}$, where $b$ is the
  semi-minor axis and $\epsilon$ is the ellipticity. In particular,
  given the observed ellipticity $\epsilon=0.3$, the circularized
  S\'ersic radius $\RS=14.5\arcmin$ corresponds to an elliptical
  radius $17.3\arcmin$.} radius is best fit by a \citet{Ser68} profile
\begin{equation}
I(R)=\Izero\exp\left[-\left(\frac{R}{\RS}\right)^{1/m}\right]
\label{eq:ser}
\end{equation}
with $\RS =14.5\arcmin$ ($\simeq 0.58$ kpc, at a distance
of 138 kpc) and $m=0.71$. For these values of $\RS$ and $m$ the
projected circularized radius containing half of the total number of
stars is $\simeq 0.62\kpc$.

The stellar mass of Fornax was estimated by \citet{deBoer12} who
performed a star formation history determination over a wide area
using what are so far the deepest available color-magnitude diagrams
over a large spatial region for this galaxy. The resulting
  stellar mass is $4.3\times 10^7\Msun$ within a projected
  circularized radius $0.66$ deg (1.6 kpc). The observed LOS velocity
dispersion profile was derived by M. Breddels from $\sim$2900
individual probable Fornax members following the approach outlined in
BH13, but binning stars in elliptical annuli\footnote{The LOS velocity
  dispersion profile published in BH13 differs from the one used here,
  because BH13 binned the data in circular annuli.}.

A central information to our analysis is the one provided by the
systemic heliocentric velocity and proper motion of Fornax, which
determines the orbit of the object around the MW and allows us to
understand how important tidal effects may have been in the evolution
of this galaxy.  In the literature there is good consistency among the
heliocentric LOS systemic velocity ($v_{\rm sys, h}$) values derived
from different spectroscopic data-sets of individual stars, even
though the samples have widely different sizes and were gathered at
various observing facilities (e.g. $v_{\rm sys, h}= 53 \pm 1.8$,
$54.1\pm 0.5$, $53.3\pm0.8 \kms$ from Mateo et al. 1991, Battaglia et
al. 2006, Walker et al. 2006, respectively).

There are several direct estimates of the systemic proper motion of
Fornax from astrometric measurements (Dinescu et al. 2004;
\citealt{Pia07}, hereafter P07; \citealt{Men11}), which are somewhat
discrepant (see table~6 of \citealt{Men11}). For example, the current
range of most likely apogalacticon estimates is very wide, from 152
kpc (P07) reaching up to the 965$\pm$356 kpc as from the work of
M{\'e}ndez et al., in which case Fornax would be at its first pericentric
passage. However, all measurements consistently suggests that Fornax
is unlikely to have passed close to the MW centre, as the most likely
perigalacticons range from 118 kpc (P07) to 148 kpc (\citealt{Men11}).
As we wish to assess the impact of tidal disturbances, we adopt the
orbital parameters from P07, which yield the tightest orbit among the
direct proper motion measurements. This determination is also in
very good agreement with indirect proper motion determinations from stellar
redshifts \citep{2008ApJ...688L..75W}. 

The values of Fornax's
observational parameters adopted in this paper are summarized in
Table~\ref{tab:prop}.

\begin{table}
\centering
\caption{Values of Fornax's observational parameters adopted in this work.
The position angle, P.A., is defined as the angle between the North
and the projected major axis of the galaxy measured anti-clockwise.
The systemic velocity $v_{\rm sys, h}$ refers to the heliocentric
reference frame, while the radial ($\vr$) and tangential ($\vt$)
velocity components are in the Galactocentric system. The quoted radii
are circularized projected radii. References: 1) \citet{Mat98}; 2)
\citet{Bat06}; 3) \citet{deBoer12}; 4) \citet{Sav00}; 5)
\citet{Pia07}.}
\label{tab:prop}
\centering
\begin{tabular}{lcc}
\hline
(l,b) & (237.1\degr, $-$65.7\degr) & 1 \\
$\epsilon$ & 0.3$\pm$0.01  & 2 \\
P.A. & $46.8^{\circ} \pm 1.6^{\circ}$ & 2 \\
$R_S$ & 14.5 arcmin &  2 \\
$m$   & 0.71 &  2 \\
$\Mstar$(R $<$ 1.6 kpc) & 4.3 $\times 10^7$ \sm & 3\\
$d$ & 138$\pm$8 kpc & 4 \\
$v_{\rm sys, h}$ & $54.1\pm 0.5\kms$ & 2\\
$\vr$ & $-31.8 \pm 1.7\kms$ & 5\\
$\vt$ & $196 \pm 29 \kms$ & 5\\
\hline
\end{tabular}
\end{table}

\section{Set-up of the $N$-body simulations}\label{sec:setup}

In order to study the effect of tides on the internal structure and
kinematics of Fornax we ran a set $N$-body simulations of a
Fornax-like object in orbit around the MW. In the simulations the MW
is represented as a fixed gravitational potential: in particular we
adopt the Galactic potential of \citet{Joh95}, which consists
of a Miyamoto \& Nagai (1975) disc
\begin{equation}
\phi_{\rm disc}=-\frac{G~M_{\rm disc}}{\sqrt{R^{2}+(a+\sqrt{z^{2}+b^{2}})^{2}}},
\label{eq:potMW1}
\end{equation}
a spherical Hernquist (1990) bulge
\begin{equation}
\phi_{\rm bulge}=-\frac{G~M_{\rm bulge}}{r+c},
\label{eq:potMW2}
\end{equation}
and a logarithmic DM halo
\begin{equation}
\phi_{\rm halo}=v_{\rm halo}^{2}~\ln(r^{2}+d_{\rm halo}^{2}),
\label{eq:potMW3}
\end{equation} 
where $r$ is the Galactocentric radial spherical coordinate, and $R$ and
$z$ are the Galactocentric radial and vertical cylindrical
coordinates.  We adopt $M_{\rm disc}=10^{11}~M_{\odot}$, $M_{\rm
  bulge}=3.4\times10^{10}~M_{\odot}$, $v_{\rm halo}=128\kms$,
$a=6.5\kpc$, $b=0.26\kpc$, $c=0.7\kpc$ and $d_{\rm halo}=12\kpc$.  This Galactic
model has been previously used in many other studies focused on the
determination of orbits of globular clusters (Dinescu et al. 2000;
\citealt{All08}) and by P07 to calculate the orbit of Fornax.

\begin{figure}
\includegraphics[width=0.5\textwidth]{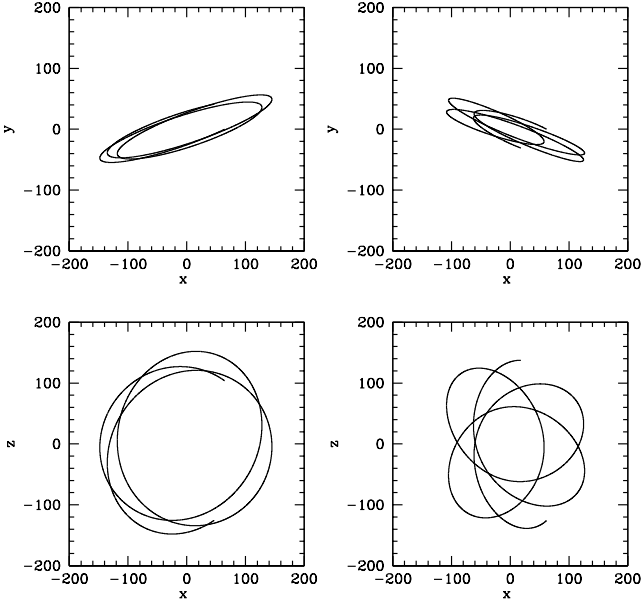}
\caption{Paths of the P07best (left-hand panels) and P07ecc
  (right-hand panels) orbits integrated in the MW potential for 12 Gyr
  up to the present-day position of Fornax. Here $x$, $y$ and $z$ are
  Cartesian Galactocentric coordinates in units of kpc ($z$ is
  orthogonal to the Galactic plane).}
\label{orbits}
\end{figure}

The phase-space coordinates of the centre of mass of the simulated
satellite at the beginning of the simulations have been derived by
taking a test particle with the present-day phase-space coordinates of
Fornax's centre and evolving it backward in time for 12 Gyr in the
adopted Galactic potential. In order to take into account the
uncertainties in the estimated proper motions, we considered two
different orbits: {\it i)} the best-fitting orbit inferred by P07
(hereafter, ``orbit P07best''); {\it ii)} the most eccentric
(i.e. with the smallest perigalacticon) orbit compatible within
$3\sigma$ with the observed radial velocity and proper motions by P07
(hereafter, ``orbit P07ecc'').  The paths of the two considered orbits
are shown in Fig.~\ref{orbits}: the orbit P07best is an almost polar
orbit, with perigalacticon $r_{\rm p}$ = 118 kpc and apogalacticon
$r_{\rm a}$ = 152 kpc, yielding an orbital eccentricity $e =(r_{\rm
  a}-r_{\rm p})/(r_{\rm a}+r_{\rm p}) \simeq 0.13$; the P07ecc
orbit has eccentricity $e\simeq 0.4$ and perigalactic radius $r_{\rm
  p}=61$ kpc. Over the considered 12 Gyr the dwarf galaxy goes through
three pericentric passages in orbit P07best and five pericentric
passages in orbit P07ecc.

In the initial conditions of the simulations Fornax is represented
with particles as a two-component, spherically symmetric system: a
stellar component embedded in a DM halo. This is an approximation,
since Fornax's stellar component is observed to be mildly flattened,
with projected ellipticity $\sim0.3$ (in Section~\ref{sec:flattened}
we will present a simulation in which we attempt to reproduce also the
observed ellipticity).  The initial structure of the satellite has
been chosen so that at the end of the simulation (after 12 Gyr) the
satellite has projected stellar-density and LOS velocity-dispersion
profiles similar to those observed for Fornax (see
Section~\ref{sec:obs}). Of course, the problem of matching the final
snapshot of the simulations with the observations is degenerate and
requires some assumptions to limit the number of free parameters.  In
particular we assumed that the initial velocity distributions of stars
and DM in the satellite are isotropic, and we also fixed the
functional form of the initial stellar and DM density profiles.

The initial stellar component of the dwarf is a spherical distribution
with density profile \citep{Lim99}
\begin{equation}
\rhostar=\rho_{0,*}\left(\frac{r}{r_{\rm
      c,*}}\right)^{-p}e^{-\left(\frac{r}{r_{\rm
      c,*}}\right)^{\nu}},
\label{eq:serdeproj}
\end{equation}
which for $\nu=1/m$ and $p=1-0.6097\nu+ 0.05463\nu^2$ is a good
approximation of the deprojection of the \citet{Ser68} profile
(equation~\ref{eq:ser}) with index $m$ and characteristic radius
$r_{\rm c,*}=\RS$. The density profile of the DM halo is not strongly
constrained by the available observations: in particular it is
uncertain whether the central DM density profile is cuspy or cored.
Though cuspy halos are found in cosmological DM-only simulations,
cored halos are generally expected in theoretical models of dSph
formation that account for the effect of baryons
\citep[e.g.][]{Nav96,Mo04,Mas06,Del09,Dic14,Pon14,Nip15}. Moreover,
cored halos tend to reproduce the observations better than cuspy halos
(e.g. \citealt{Wal11}; \citealt{Jar12};
\citealt{Amo13}), though this finding is still debated (see
e.g. BH13).  Since previous works found that the stellar component of
simulated galaxies within cuspy DM halos is more resilient to tides
than in cored DM halos (e.g. \citealt{Pen10}), the assumption of a
cored DM profile should enhance the pace at which our satellite is
affected by tidal disturbances.  Here we represent the initial DM halo
of the dwarf with a density profile
\begin{equation}
\rhoDM(r)=\rho_{\rm 0,DM}\left(1+\frac{r}{\rcDM}\right)^{-3}e^{-\left(\frac{r}{\rtDM}\right)^{2}},
\label{eq:rhodm}
\end{equation}
which can be thought of as a NFW-like profile \citep{Nav95} with scale
radius $\rcDM$, but with no central cusp ($\d \ln \rho/\d \ln r=0$ at
$r=0$) and truncated exponentially at radii larger than $\rtDM$.

In all the simulations we fixed $m=0.71$ (the observed best-fitting
S\'ersic index; see Section~\ref{sec:obs}),
$\rcDM=3.5\kpc$\footnote{The choice of a relatively large value
    of $\rcDM$ is needed to reproduce the flat shape of the LOS
    velocity-dispersion profile observed in Fornax, under our
    assumption of isotropic stellar velocity distribution. A similar
    LOS velocity-dispersion profile can be obtained with larger values
    of $\rcDM$, combined with higher values of DM mass, but in
    practice such haloes would extend far beyond the tidal limit and
    lose a large fraction of their particles at the beginning of the
    simulation. For this reason, and to limit the number of free
    parameters, we decided to keep the value fo $\rcDM$ fixed.} and
$\rtDM=6 \kpc$, while the initial stellar scale radius $\rcstar$,
total stellar mass $\Mstar =4\pi\int_0^{\infty}r^2\rhostar(r) \d r$
and total DM mass $\MDM=4\pi\int_0^{\infty}r^2\rhoDM(r) \d r$ are left
as free parameters.

The values of $\rcstar$, $\Mstar$ and $\MDM$ are
tuned in each simulation to reproduce in the last snapshot the
projected surface-density and LOS velocity-dispersion profiles of
Fornax, and its observationally estimated stellar mass. For this purpose,
we used the following iterative procedure.
\begin{enumerate} 
\item{We first adopt guess values of $\rcstar$, $\Mstar$ and $\MDM$
  that give good fits to the present-day projected surface-density
  and LOS velocity-dispersion profiles of Fornax.}
\item{When a simulation with these values of the parameters is run,
  its final snapshot is projected on the plane of the sky and used to
  measure the stellar mass within a projected radius of 1.6 kpc
  ($M_{*,1.6}^{\rm sim}$), the half-light radius ($R_{\rm hl}^{\rm sim}$)
  and the LOS velocity dispersion within the half-light radius
  ($\sigma_{\rm LOS,hl}^{\rm sim}$).}
\item{New values of $\rcstar$, $\Mstar$ and $\MDM$ (here indicated
  with primes) are then calculated as follows:
\begin{eqnarray*}
\rcstar'&=&\rcstar~(R_{\rm hl}^{\rm obs}/R_{\rm hl}^{\rm sim})^{0.5},\nonumber\\
\Mstar'&=&\Mstar~(M_{*,1.6}^{\rm obs}/M_{*,1.6}^{\rm sim})^{0.5},\nonumber\\
\MDM'&=&\MDM~(\sigma_{\rm LOS,hl}^{\rm obs}/\sigma_{\rm LOS,hl}^{\rm sim}),\nonumber
\end{eqnarray*}
where $M_{*,1.6}^{\rm obs}=4.3\times 10^7 M_{\odot}$, $R_{\rm hl}^{\rm
  obs}=0.6 \kpc$ and $\sigma_{\rm LOS,hl}^{\rm obs}=12\kms$ are the
observational estimates for Fornax (see Section \ref{sec:obs}).}
\end{enumerate}

Steps (ii) and (iii) are repeated until convergence (defined by the
condition that the three free parameters change by less than 10\%). For
the simulations presented here, the above scheme converged after 2-3
iterations.  It must be noted that our iterative procedure is not as
robust as other more complex algorithms developed for this purpose
(see e.g. Ural et al. 2015) and could potentially fail to converge if
significant changes in the structure of the satellite
occurred. However, given the relatively wide orbit of Fornax, in all
our simulations the three parameters change by small amounts during
the entire satellite evolution (see Section \ref{sec:results}), so we
believe that the simple procedure adopted here provides reliable
results in the present context.

The $N$-body simulations have been performed with the collisionless
code \FVFPS \citep[][]{Lon03,Nip03}, which has been specifically
modified to include the fixed external potential of the MW
(equations~\ref{eq:potMW1}-\ref{eq:potMW3}).  We adopted the following
values of the code parameters: minimum value of the opening parameter
$\theta_{\rm min}=0.5$, softening parameter $\varepsilon = 10$ pc, and
(constant) time-step $\Delta t=3.2\times10^5\yr$. The initial
equilibrium two-component satellite system is generated assuming
isotropic velocity distribution: given the density distribution of
each component and the corresponding total gravitational potential,
the particles are assigned phase-space coordinates $x$, $y$, $z$,
$\vx$, $\vy$, $\vz$ (centred in the dwarf galaxy's centre of mass)
using the ergodic distribution function obtained with Eddington's
inversion \citep{Bin08}.  The parameters of the simulations, including
the number of stellar ($N_*$) and DM ($N_{\rm DM}$) particles, are
given in Table~\ref{tab:sim}.

\begin{table*}
\centering
\caption{Initial parameters of the simulations. $\rcstar$: scale
  radius of the stellar distribution; $\Mstar$: stellar mass; $\MDM$:
  DM mass; $\Mtot$: total mass; $N_{*}$: number of stellar particles;
  $N_{\rm DM}$: number of DM particles; $N_{\rm tot}$: total number of
  particles; $e$: orbit eccentricity.  {\it Notes.} For simulation
  P07best$_{\rm flat}$, in which the stellar component is initially
  flattened, $\rcstar$ refers to the spherical distribution used in
  the set-up of the initial conditions (see
  Section~\ref{sec:flattened}).  In simulations P07best$_{\rm MFL}$
  and P07ecc$_{\rm MFL}$ there is no distinction between stars and DM,
  so $\rcstar$ is the scale radius of the total mass distribution. }
\label{tab:sim}
\begin{tabular}{lcccccccr}
\hline
Name & $\rcstar$ & $\Mstar$ & $\MDM$ & $\Mtot$ & $N_{*}$  & $N_{\rm DM}$ & $N_{\rm tot}$ & $e$\\
     & kpc       & $\Msun$ & $\Msun$ & $\Msun$ &  &  & & \\
\hline
P07best    & 0.581 & $5\times10^{7}$   & $3\times10^{9}$ & $3.05\times10^{9}$ & 51200  & 1536000   & 1587200   & 0.13\\
P07ecc     & 0.581 & $5\times10^{7}$  & $3\times10^{9}$ &   $3.05\times10^{9}$ & 51200  & 1536000  & 1587200   & 0.40\\
P07ecc$_{\rm sh}$ & 0.465 & $5\times10^{7}$  & $4.2\times10^{9}$ & $4.25\times10^{9}$ & 51200  & 2150400 &  2201600 & 0.40\\
P07best$_{\rm MFL}$  & 0.581 & - & - & $1.5\times10^{8}$ & - & - &     51200       & 0.13\\
P07ecc$_{\rm MFL}$   & 0.581 & - & - & $1.5\times10^{8}$ & - & - &     51200        & 0.40\\
P07best$_{\rm flat}$   & 0.581 & $5\times10^{7}$ & $2.6\times10^{9}$ & $2.65\times10^{9}$ & 51200 & 1331200 &    1382400        & 0.13\\
\hline
\end{tabular}
{\footnotesize}
\end{table*}

\section{Results} \label{sec:results}

\subsection{Evolution of the satellite's stellar and dark-matter components}

We present here the results of three simulations (P07best, P07ecc and
P07ecc$_{\rm sh}$) differing in either the satellite orbit or the
initial stellar and DM density distributions (see
Table~\ref{tab:sim}).  Simulations P07best and P07ecc have the same
initial DM and stellar distribution and differ only in the orbital
parameters ($e=0.13$ and $e=0.4$, respectively).  In simulation
P07ecc$_{\rm sh}$ the satellite is placed on the more eccentric orbit
(as in simulation P07ecc), but the stellar component starts with a
more centrally concentrated spatial distribution than in simulation
P07ecc, and a DM mass higher by a factor 1.4. In all these simulations
the total stellar mass is $\Mstar=5\times10^7\Msun$. The simulations
are run for 12 Gyr: the final snapshot corresponds to the present day
and can be compared with the observational data of Fornax.

\begin{figure*}
\includegraphics[width=\textwidth]{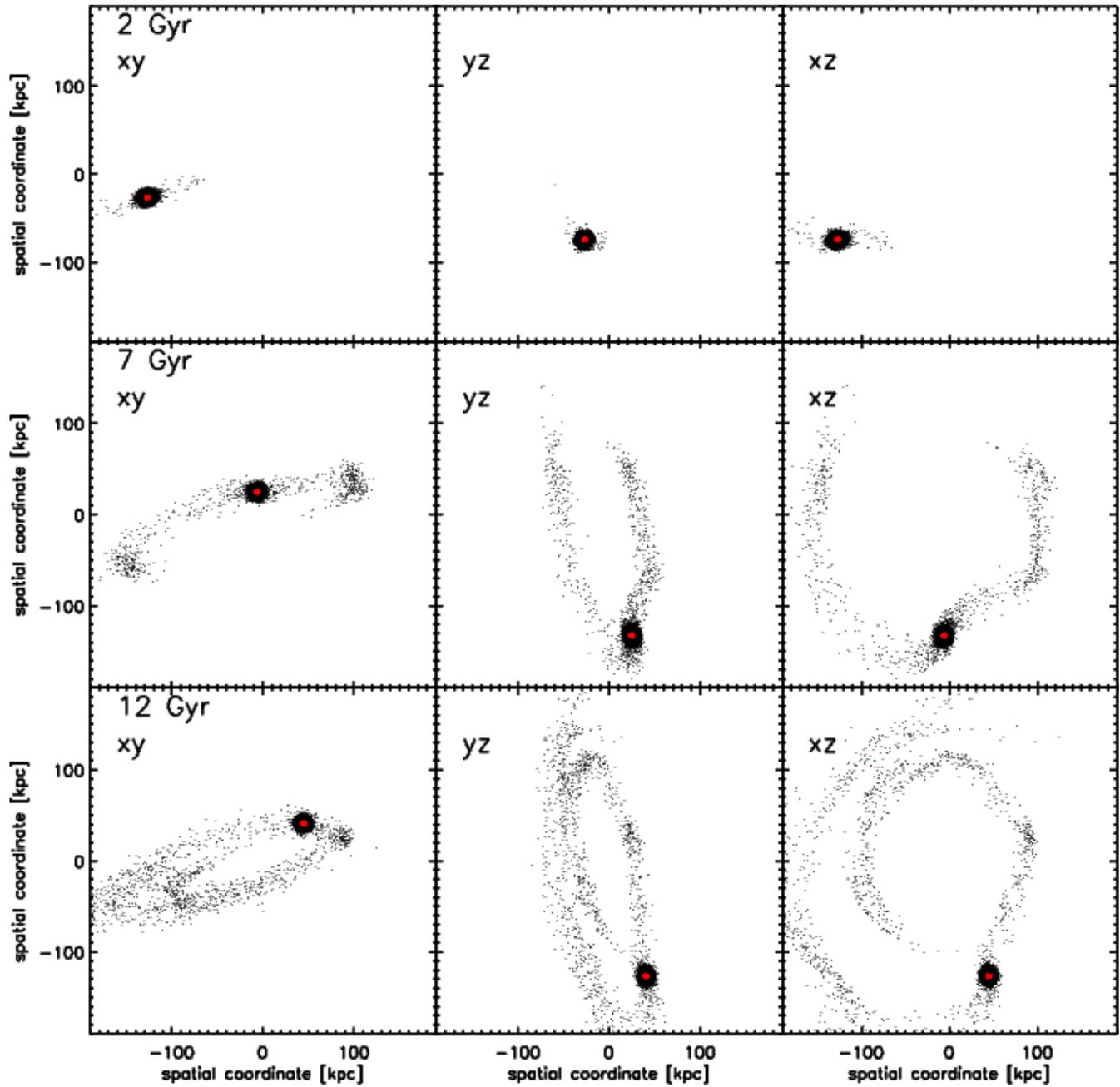}
\caption{Distribution of DM (black) and stellar (red) particles of the
  satellite galaxy as seen in the $x$-$y$, $y$-$z$, and $x$-$z$ planes
  (from left to right) for simulation P07best, after 2, 7 and 12 Gyr
  of evolution (from top to bottom).  Here $x$, $y$ and $z$ are
  Cartesian Galactocentric coordinates in units of kpc ($z$ is
  orthogonal to the Galactic plane).  For convenience we plot one
  every thirty DM particles. }
\label{fig:xyz_p07}
\end{figure*}

\begin{figure*}
\includegraphics[width=\textwidth]{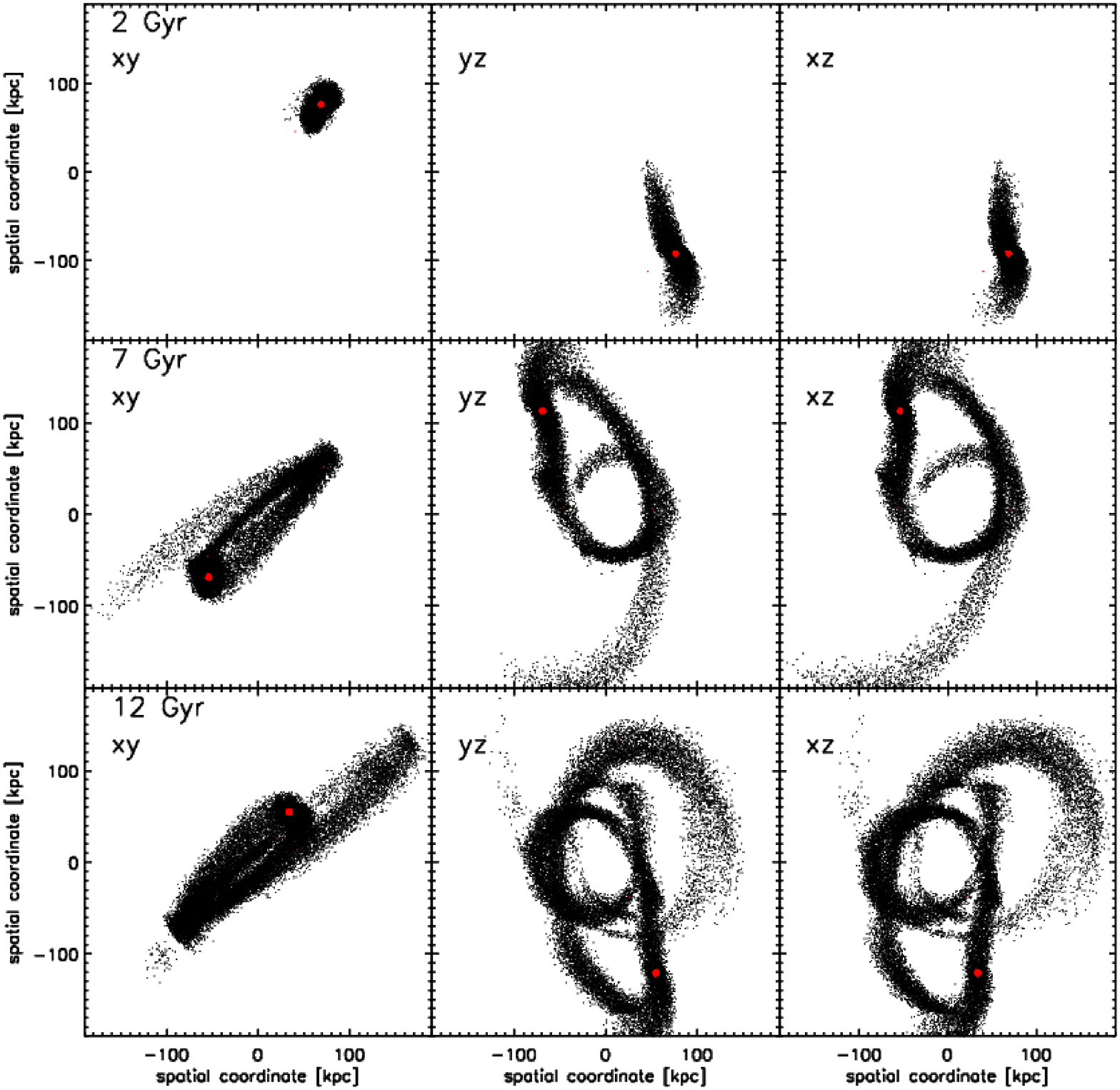}
\caption{Same as  Fig.~\ref{fig:xyz_p07}, but for simulation P07ecc. }
\label{fig:xyz_p07ecc}
\end{figure*}

\begin{figure*}
\includegraphics[width=\textwidth]{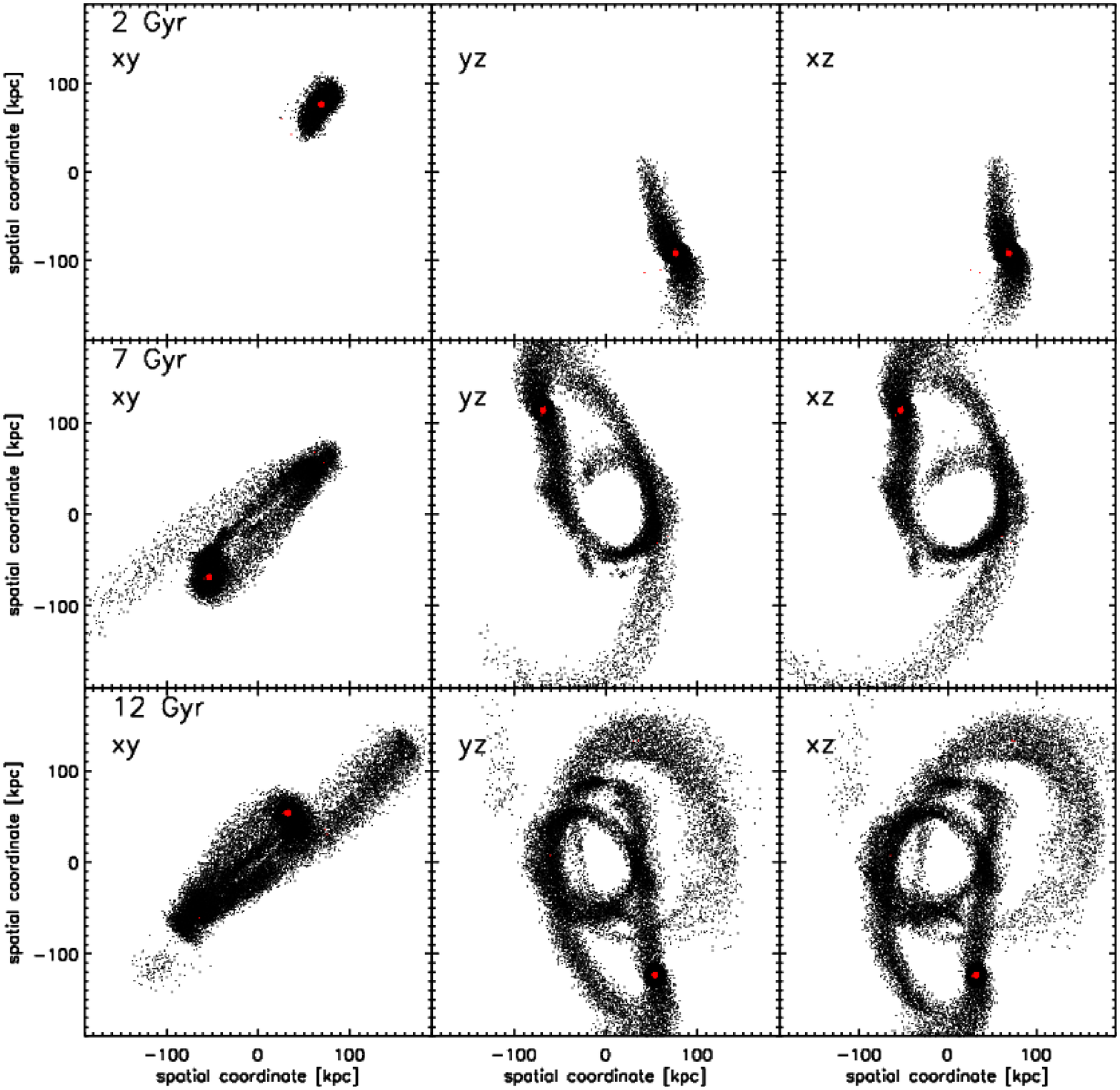}
\caption{Same as Fig.~\ref{fig:xyz_p07}, but for simulation
  P07ecc$_{\rm sh}$}
\label{fig:xyz_p07ecc_sh}
\end{figure*}

Figures~\ref{fig:xyz_p07}-\ref{fig:xyz_p07ecc_sh} show the spatial
distribution of DM and stellar particles after 2, 7 and 12 Gyr for the
simulations P07best, P07ecc and P07ecc$_{\rm sh}$.  In all cases the
DM component develops tidal tails: the fraction of particles contained
in the tidal tails is clearly higher in the simulations on the more
eccentric orbit. Tidal tails are instead absent in the stellar
component in all these simulations.  This can be qualitatively
understood considering that along the P07best orbit the minimum
(pericentric) tidal radius, calculated according to the prescriptions
of \citet[][see their appendix A]{All06} and assuming a mass for
Fornax of $3 \times 10^{9} M_{\odot}$ (see Table \ref{tab:sim}), is
$r_{\rm t,min}=13.16$ kpc, which is large compared to the
characteristic size of the stellar component of Fornax (in the initial
conditions of the simulations about 99\% of the stellar mass is
contained within $2\kpc$). The more eccentric orbit P07ecc has a
minimum tidal radius $r_{\rm t,min}=7.36$ kpc, which is significantly
smaller than that of orbit P07best, but still larger than the radius
containing most of the stellar mass.  So, during the satellite orbit,
stars are always contained within the instantaneous tidal radius and
are not expected to be significantly affected by tides.  Taken at face
value, the above estimates of the tidal radii are relative large also
with respect to the characteristic size of the DM distribution (in the
initial conditions about $90\%$ of the DM mass is contained within
$r=7.5\kpc$), so the above simple analysis would predict moderate
tidal effects also for the halo.  In fact, the $N$-body simulations
show that the DM halo develops evident tidal tails and lose a
significant fraction of its mass. The stellar component is not tidally
stripped, but, at least on the more eccentric orbit, its internal
structure and kinematics evolve due to the variation of the
gravitational potential induced by the tidal stripping of the halo.

\begin{figure*}
\includegraphics[width=\textwidth]{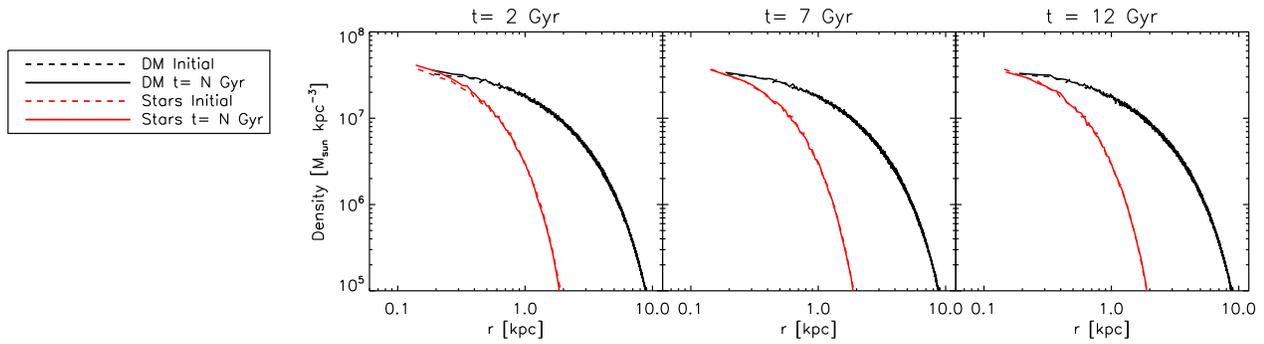}
\caption{Angle-averaged density profiles of DM (black) and stellar
  (red) particles for three snapshots, at 2, 7 and 12 Gyr, for the 
  simulation P07best (solid lines) as compared to the 
  initial conditions (dashed lines). }
\label{fig:rho_p07}
\end{figure*}

\begin{figure*}
\includegraphics[width=\textwidth]{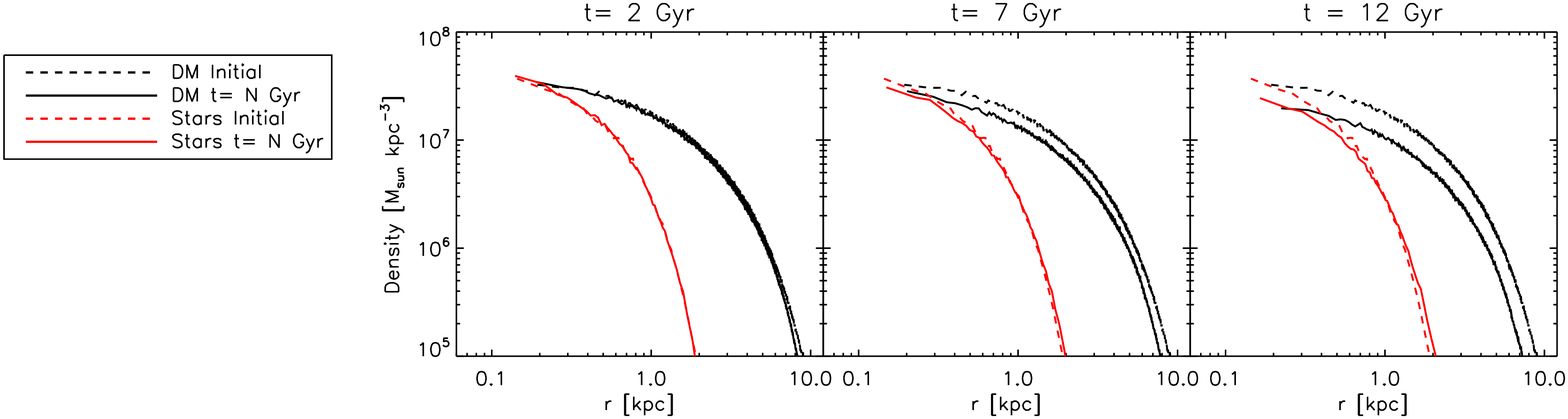}
\caption{Same as Fig.~\ref{fig:rho_p07}, but for simulation P07ecc. }
\label{fig:rho_p07ecc}
\end{figure*}

\begin{figure*}
\includegraphics[width=\textwidth]{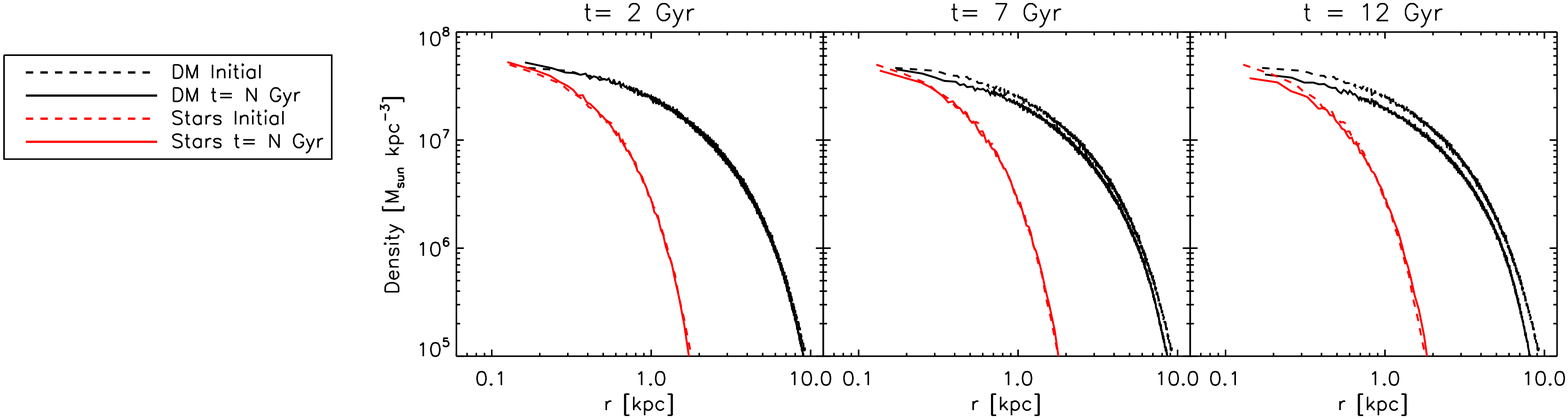}
\caption{Same as Fig.~\ref{fig:rho_p07}, but for simulation P07ecc$_{\rm sh}$.}
\label{fig:rho_p07ecc_sh}
\end{figure*}

Figures~\ref{fig:rho_p07}-\ref{fig:rho_p07ecc_sh} show the intrinsic
angle-averaged stellar and DM density profiles of the satellite in
simulations P07best, p07ecc and p07ecc$_{\rm sh}$ at $t=2$, 7 and 12
Gyr, compared to the initial
ones. Figures~\ref{fig:obs_p07}-\ref{fig:obs_p07ecc_sh} show, for the
same simulations, the initial ($t=0$) and final ($t=12$ Gyr) projected
number-density and LOS velocity-dispersion profiles compared to the
observations.  The final projected properties of the satellite are
computed by assuming that the system is observed from the Sun.  Given
that we consider an axisymmetric MW potential, we have the freedom to
fix the position of the Sun a posteriori in a circle of radius $8.5$
kpc lying in the Galactic plane and centred in the Galactic
centre. The Sun position is chosen in each simulation to obtain the
best possible match with the heliocentric centre-of-mass position and
systemic velocity of Fornax which are in all the simulations presented here
reproduced to within 5\%.

The final properties of the dwarf galaxies in the simulations are
summarized in Table~\ref{tab:final}. In simulation P07best neither the
DM nor the stellar distributions evolve substantially.  At the end of
the simulation $97.5\%$ of the DM particles and virtually all the
stellar particles are bound to the satellite.  At the end of the
simulation both the projected number-density and LOS
velocity-dispersion profiles of the stars are very similar to the
initial ones and compare well to the observations.  A convenient
reference radius to analyze the properties of the satellite is 1.6
kpc, which contains about $90\%$ of the stellar mass in the initial
conditions, besides corresponding approximately to the outermost point
of the observed LOS velocity dispersion profile of Fornax, within
which \citet{deBoer12} estimated Fornax's stellar mass.  From the
start to the end of the simulation, the DM mass within 1.6 kpc changes
little, from $2.68\times10^8$ to $2.64\times10^8\Msun$. The final
stellar mass within 1.6 kpc is $M_{*,1.6}\simeq 4.6\times 10^7\Msun$,
consistent within $7\%$ with the estimate of \citet{deBoer12}.

A satellite with the same initial conditions, placed onto the more
eccentric orbit (simulation P07ecc), experiences a significant
evolution due to tidal forces: the surface-density profile of the
stars becomes more extended and the average LOS velocity dispersion
becomes lower (see Fig.~\ref{fig:obs_p07ecc}). These changes are
mainly due to the loss of mass from the DM halo, whose density profile
is substantially altered over 12 Gyr of evolution (see
Fig.~\ref{fig:rho_p07ecc}).  At the end of the simulation the
satellite retains only $50\%$ of the DM particles (the DM mass within
1.6 kpc is 59\% of the initial value). Because of the consequent
variation of the gravitational potential, the satellite reaches an
equilibrium configuration characterized by a larger scale length and
lower velocity dispersion (the stellar component retains 92\% of its
mass within 1.6 kpc). A similar behaviour was noticed in previous
works studying the effects of tides on dwarf galaxies
(e.g. \citealt{Pen08}). Overall, the final snapshot of
simulation P07best is not a good representation of the observations:
the stellar particle surface-density profile of the simulated galaxy
is too extended with respect to the observed one, and the LOS
velocity-dispersion profile is too low (see
Fig.~\ref{fig:obs_p07ecc}).

In order to reconcile with the observations the final properties of
our simulated galaxy when placed on the more eccentric orbit, the
initial conditions had to be modified, starting with a more
concentrated stellar density profile and increasing the initial DM
mass (model P07ecc$_{\rm sh}$). As shown in
Fig.~\ref{fig:obs_p07ecc_sh}, the final snapshot of this simulation is
in good agreement with the observations.  In this case, the fraction
of bound DM particles at the end of the simulation is $71\%$ (the
final DM mass within 1.6 kpc is 79\% of the initial value).  As in
simulations P07best and P07ecc, virtually all the stellar particles
are bound to the object after 12 Gyr of evolution.  The final stellar
mass within 1.6 kpc is more than 98\% of the initial value: the object
expands slightly, with an associated small decrease in the central LOS
velocity dispersion.

The results of simulations P07best and P07ecc$_{\rm sh}$ show that the
present-day observed structure and kinematics of Fornax can be
reproduced by a system whose stellar component varies only little
during its 12 Gyr journey in the MW on observationally-motivated
orbits. In particular, the simulated systems are found to be
quasi-stationary in the last Gyr of evolution. This suggests that
Fornax can be reasonably well modeled as an isolated system in
equilibrium.  The final fraction of unbound stars in the simulations
is negligible, so the spectroscopic samples of Fornax stars should not
be contaminated significantly by tidally stripped
stars. Interestingly, baryons and DM appear to contribute similarly to
the mass budget in the central regions of Fornax, suggesting that this
galaxy may not be ``DM dominated at all radii'', as believed to be the
case for most MW dSphs.

\subsection{Jeans analysis of the final snapshots}

As a further test of the dynamical status of Fornax, we carried out an
equilibrium mass modeling analysis of the final snapshots of
simulations P07best and P07ecc$_{\rm sh}$, which were shown to match
well the observed properties of Fornax.  Should the stellar component
of the simulated systems be out of dynamical equilibrium, then we
would expect to find a discrepancy between the DM mass determined from
the equilibrium analysis and the ``true'' DM mass as measured directly
from the snapshot.

\begin{figure}
\includegraphics[width=8.8cm]{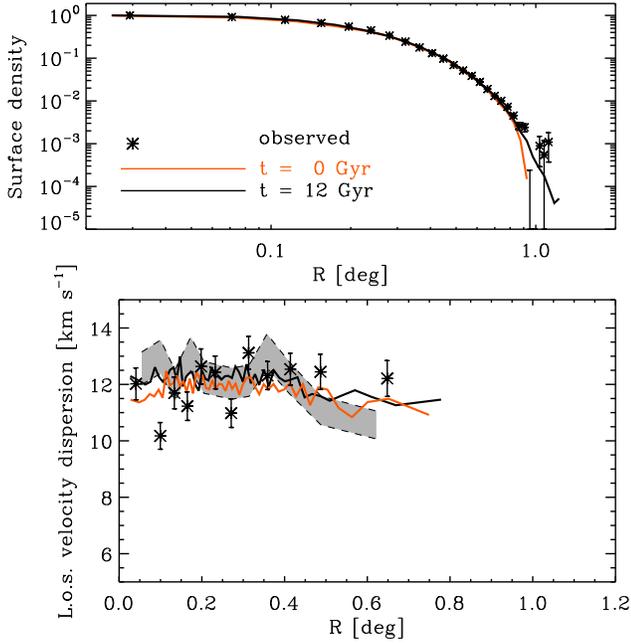}
\caption{Surface number-density (normalized to the central value; top
  panel) and LOS velocity-dispersion (bottom panel) profiles of the
  stellar component of the P07best model. The orange and black lines
  indicate the initial conditions (t=0) and the final properties
  ($t=12$ Gyr), respectively.  The observed surface number-density
  profile from Battaglia et al. (2006) and LOS velocity-dispersion
  profile from BH13 are shown as black asterisks with error bars. The
  grey band in the bottom panel refers to the LOS velocity-dispersion
  profile from the 12 Gyr snapshot when using a similar number of
  stellar particles and binning as done by BH13 for the observations
  (3000 particles, 250 per bin).}
\label{fig:obs_p07}
\end{figure}

\begin{figure}
\includegraphics[width=8.8cm]{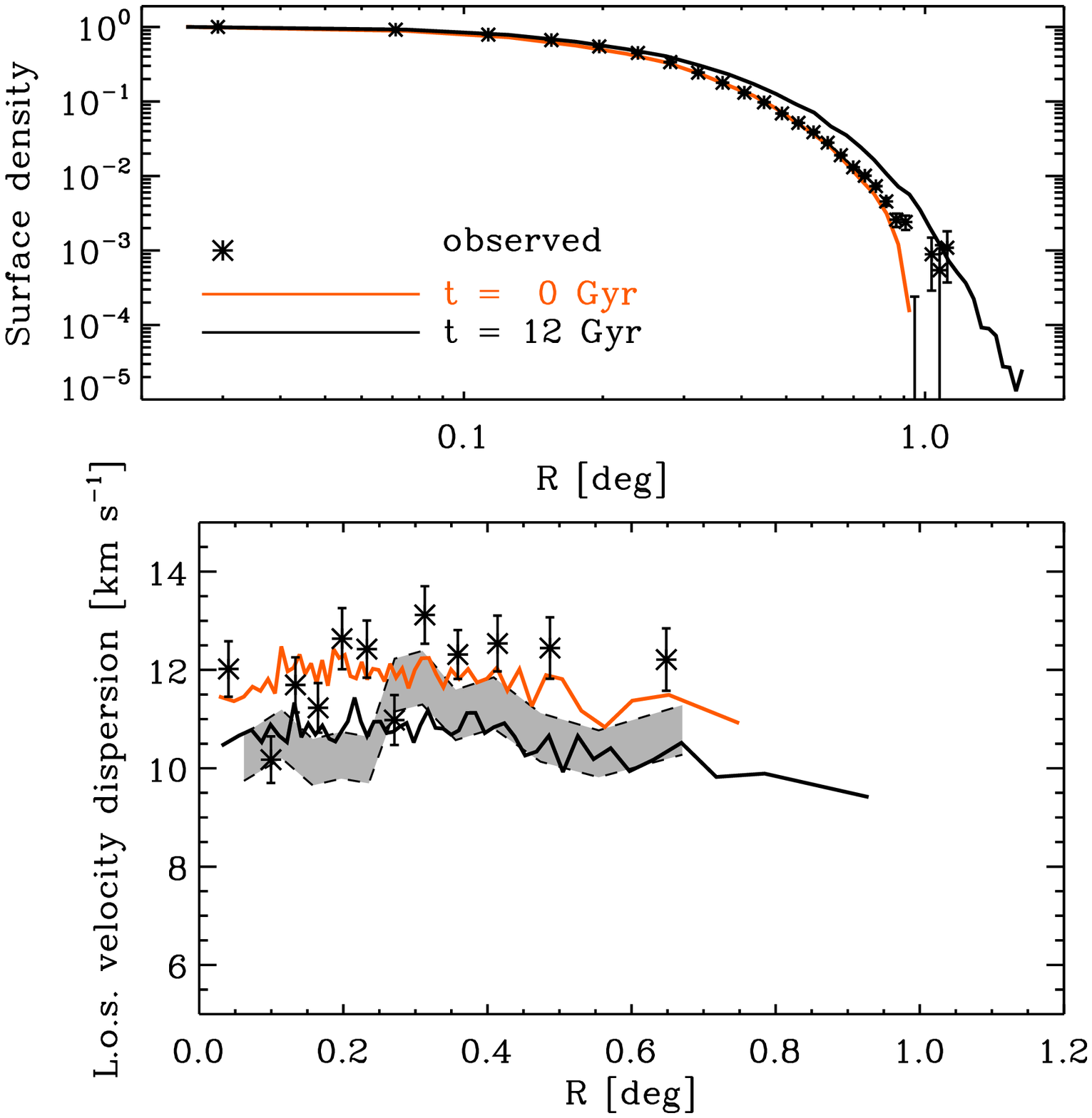}
\caption{Same as Fig.~\ref{fig:obs_p07}, but for the P07ecc model.}
\label{fig:obs_p07ecc}
\end{figure}

\begin{figure}
\includegraphics[width=8.8cm]{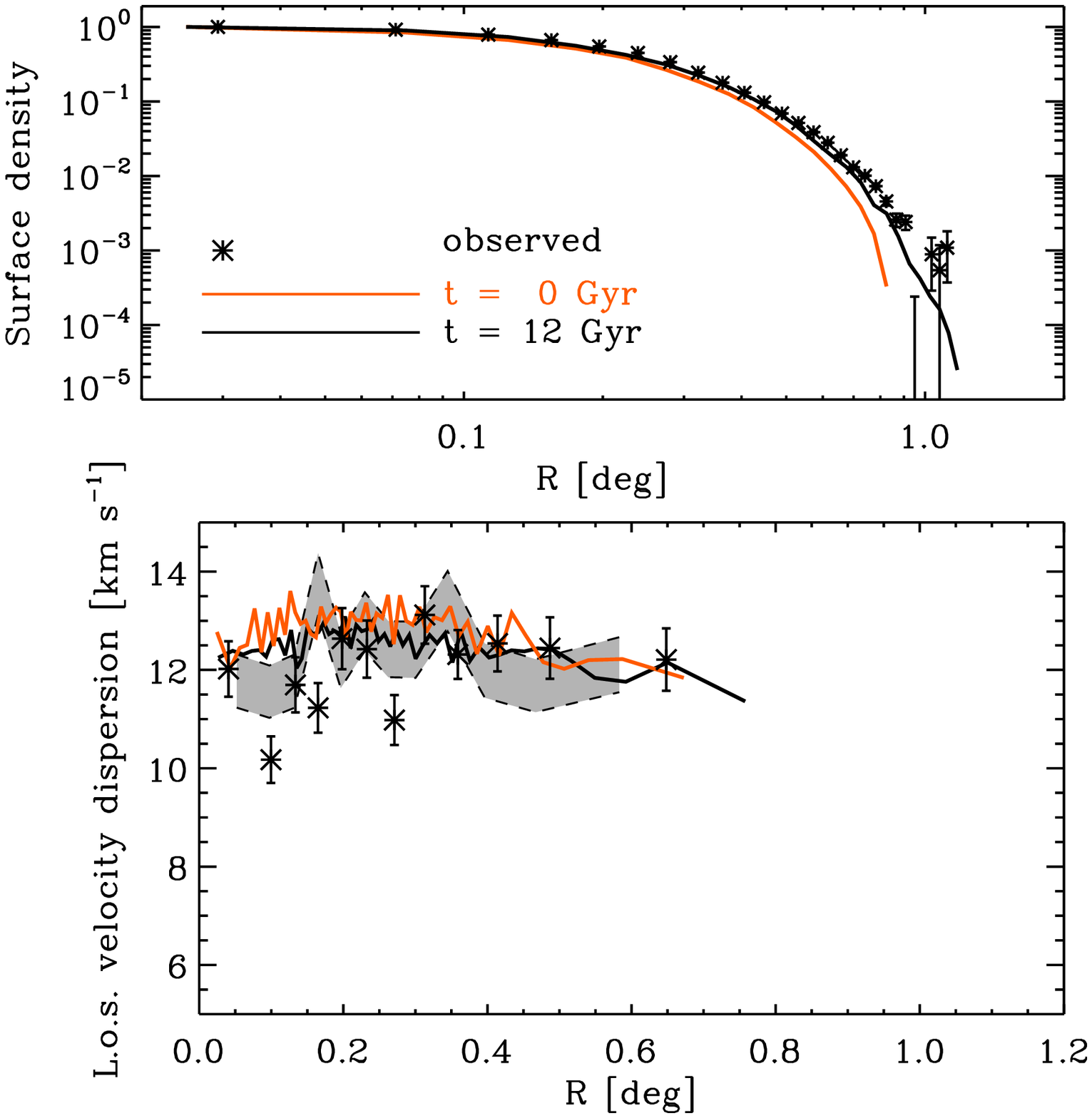}
\caption{Same as Fig.~\ref{fig:obs_p07}, but for the P07ecc$_{\rm sh}$ model.}
\label{fig:obs_p07ecc_sh}
\end{figure}

The simulated stellar component is our tracer of the gravitational
potential: given the LOS velocity-dispersion and surface
number-density profiles measured in the final snapshots of the
simulations (Figs.~\ref{fig:obs_p07} and \ref{fig:obs_p07ecc_sh}), we
wish to find the total mass distribution using the spherical Jeans
equations \citep{Bin82} under the assumption that the velocity
distribution of the tracers is isotropic.  We model the total mass
density profile as the sum of the stellar and DM density profiles.
The tracers' density and mass profiles are obtained by deprojecting
(assuming spherical symmetry) the S\'ersic profile
(equation~\ref{eq:ser}) with parameters as in Table~\ref{tab:prop} and
stellar mass $5\times 10^7$\sm.  For the DM density profile we assume
a simple model of the form
\begin{equation}
\rho_{\rm DM}(r)=\rho_{\rm 0,DM}\left(1+\frac{r}{r_{\rm c,DM}}\right)^{-4}
\end{equation} 
\citep{Deh93}, which yields the analytic mass profile
\begin{equation}
M_{\rm DM}(r)= M_{\rm DM,\infty}\left(\frac{r}{r_{\rm c,DM} + r}\right)^3,
\end{equation} 
where $M_{\rm DM,\infty} = 4 \pi \rho_{\rm 0,DM} r_{\rm c,DM}^3/3$ is
the total DM mass.

We determine the best-fitting $M_{\rm DM,\infty}$ and $r_{\rm c,DM}$
via a $\chi^2$ minimization. Though these two parameters are quite
degenerate, we find that the DM mass enclosed within 1.6 kpc ($M_{\rm
  DM,1.6}$) is not significantly affected by this degeneracy ($M_{\rm
  DM,\infty}$ is anyway a ill-constrained quantity due to the limited
spatial extent of the tracer kinematic data with respect to the extent
of the DM halo).  We obtain a remarkably good agreement between
$M_{\rm DM,1.6}$ yielded by the equilibrium Jeans analysis and the
values measured in the final snapshots of simulations P07best and
P07ecc$_{\rm sh}$, with a difference of at most 7\%.
This lends support to the fact
  that, at the present-day position of Fornax and along orbits compatible
  with literature proper motion measurements, it is
  reasonable to assume that tidal effects are not significant for DM
  determinations of this satellite galaxy within the region probed so
  far by the kinematic data.

\subsection{Mass-follows-light models}

\begin{figure*}
\includegraphics[width=\textwidth]{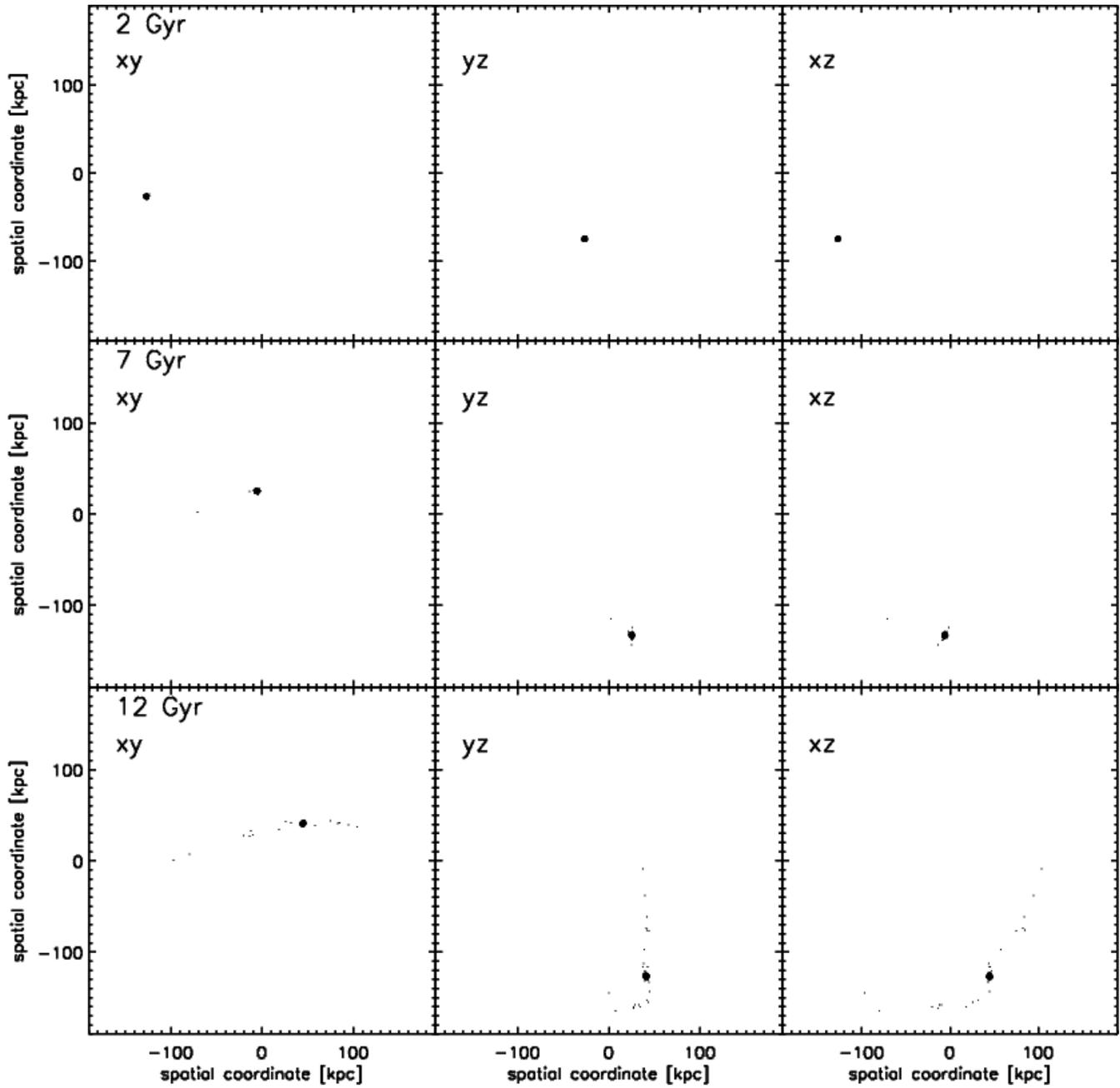}
\caption{Same as Fig.~\ref{fig:xyz_p07}, but for simulation
  P07best$_{\rm MFL}$.  Here mass follows light, so we do not
  distinguish between stellar and DM particles.}
\label{fig:xyz_p07_nodm_ml6}
\end{figure*}

\begin{figure*}
\includegraphics[width=\textwidth]{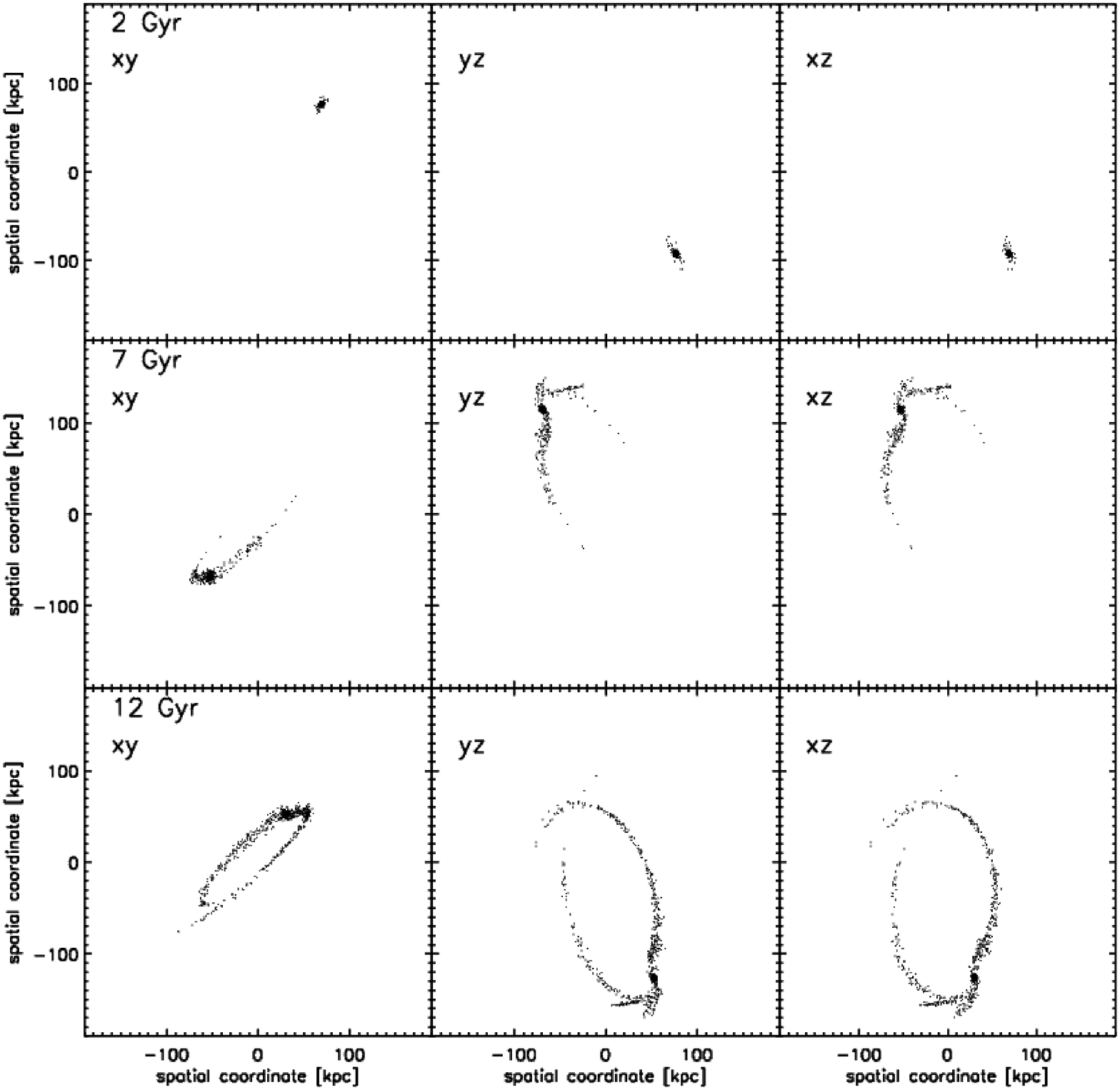}
\caption{Same as Fig.~\ref{fig:xyz_p07_nodm_ml6}, but for simulation P07ecc$_{\rm MFL}$.}
\label{fig:xyz_p07ecc_nodm_ml6}
\end{figure*}

\begin{figure*}
  \includegraphics[width=0.45\textwidth]{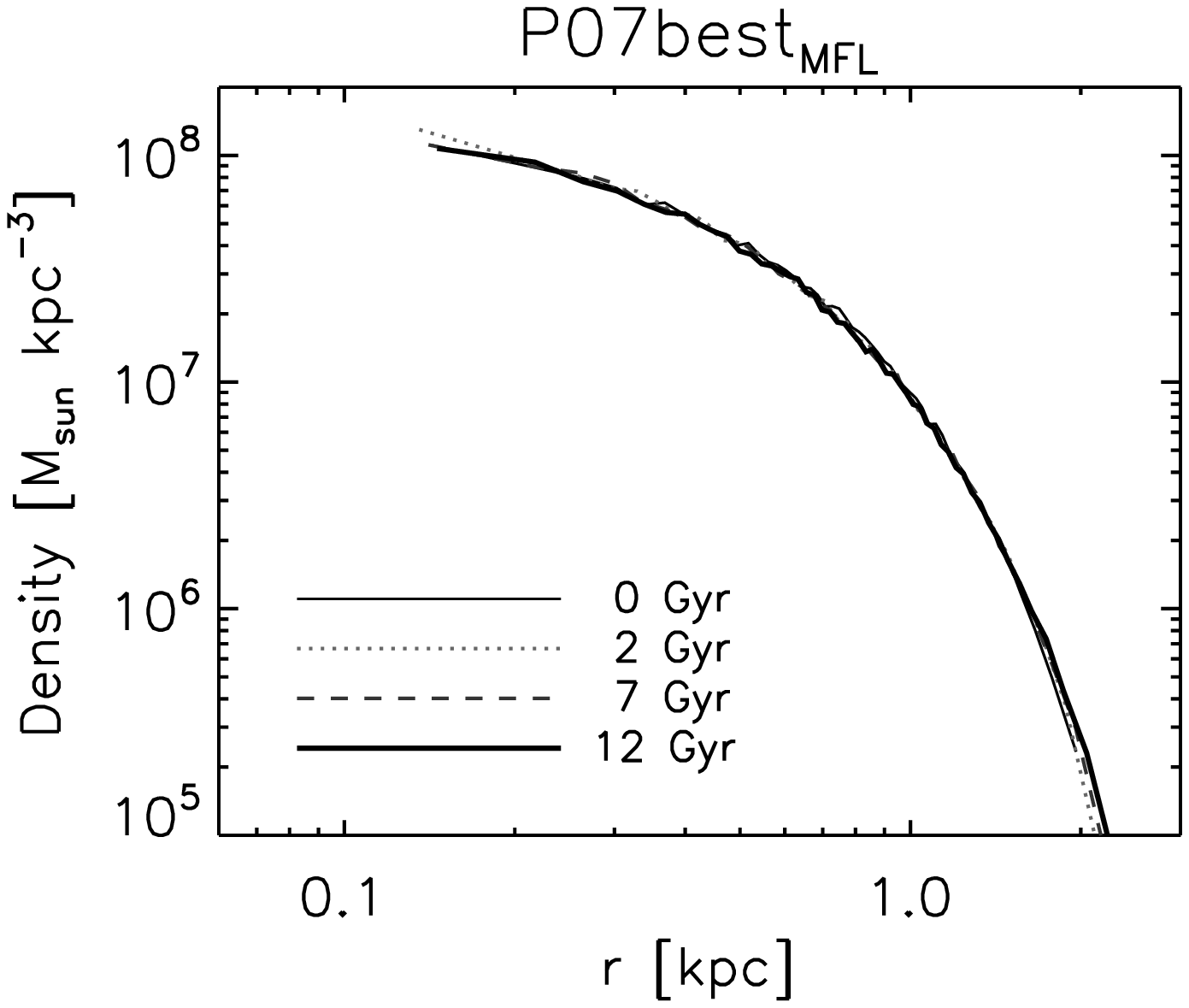}
  \includegraphics[width=0.45\textwidth]{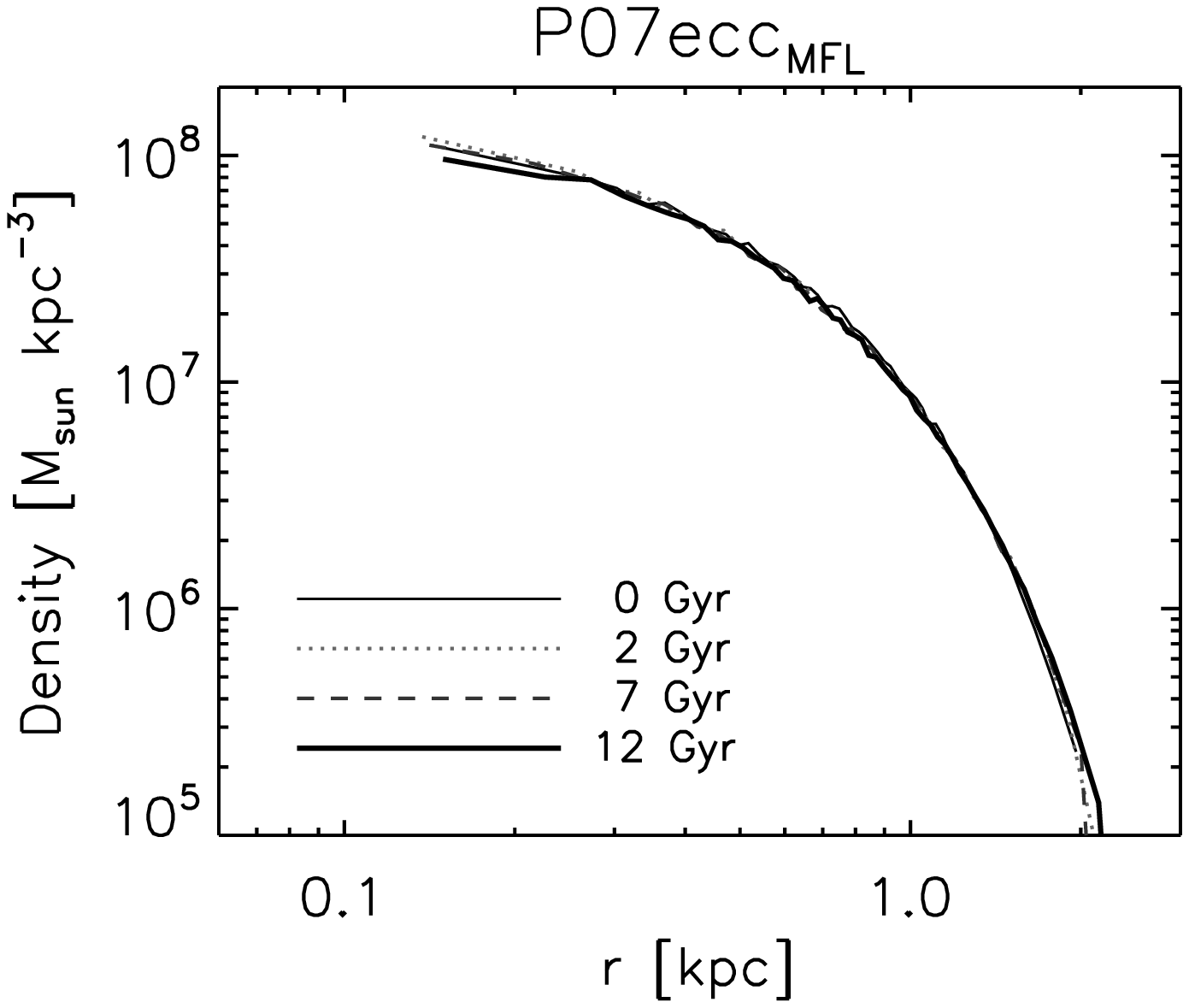}
\caption{Angle-averaged density profiles for four snapshots, at 0, 2, 7
  and 12 Gyr, for the P07best$_{\rm MFL}$ (left) and P07ecc$_{\rm
    MFL}$ (right) models. Here mass follows light, so we do not
  distinguish between stars and DM. }
\label{fig:rho_nodm_ml6}
\end{figure*}

\begin{figure*}
  \includegraphics[width=8cm]{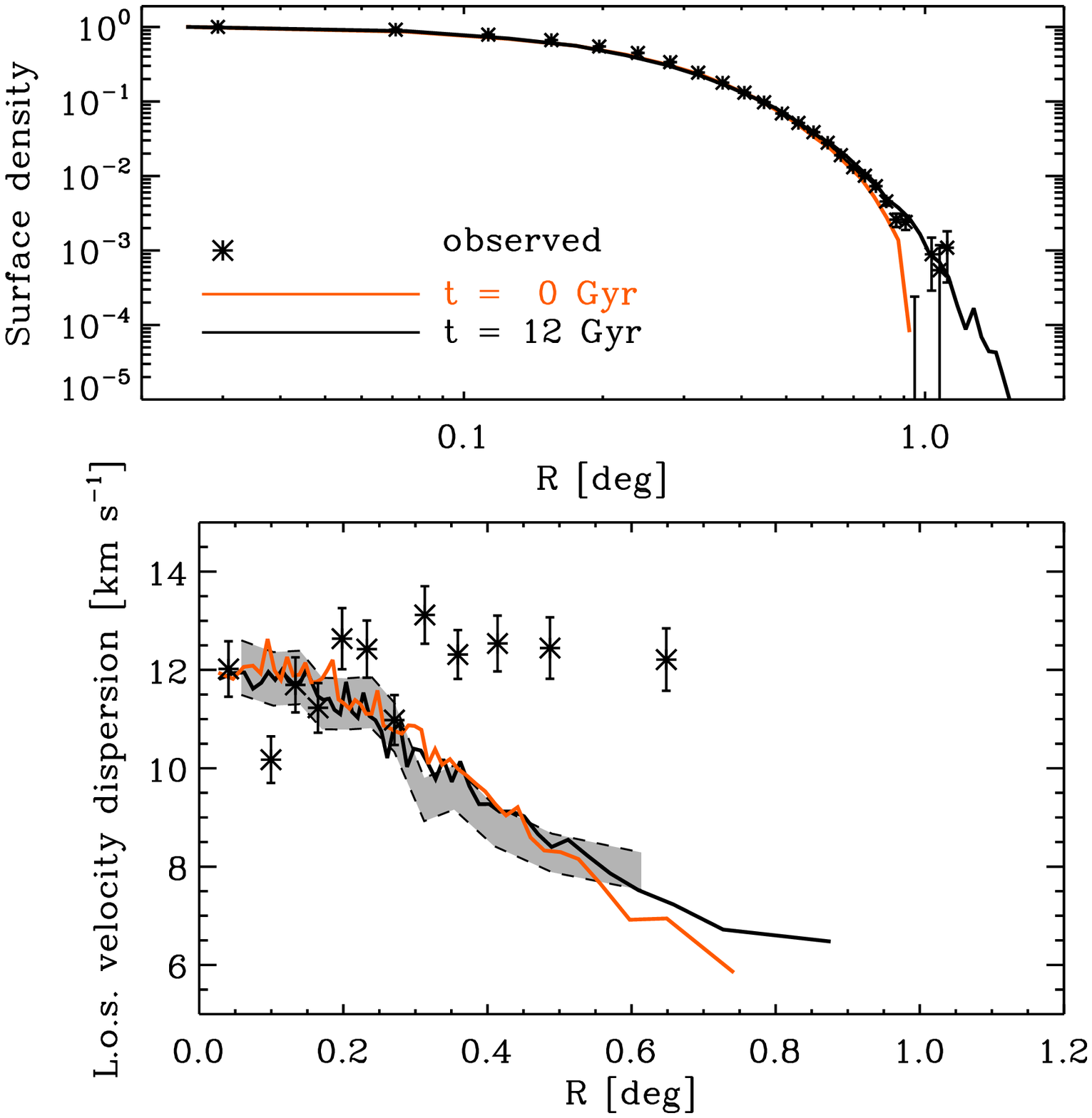}
  \includegraphics[width=8cm]{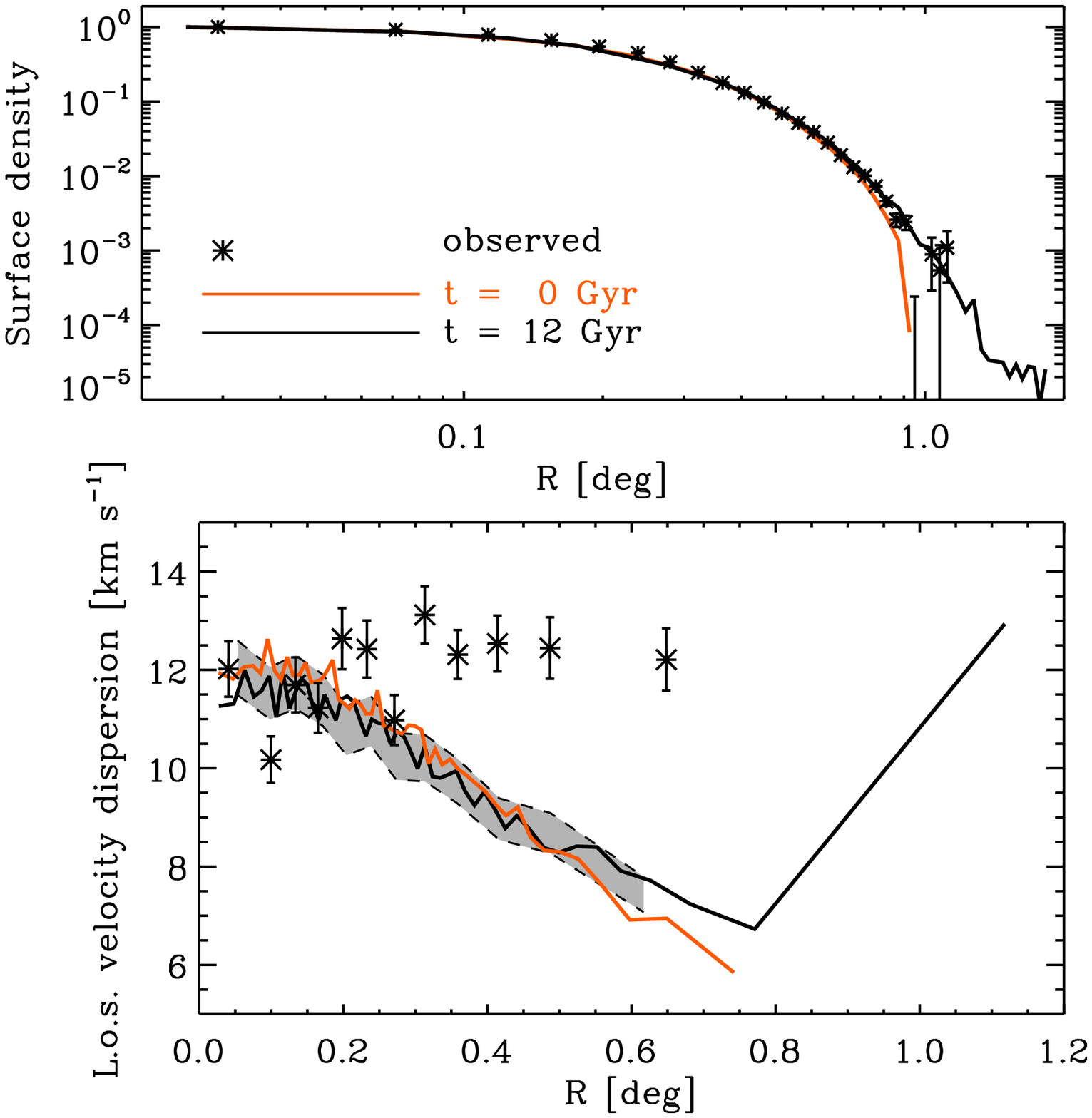}
\caption{As in Fig.~\ref{fig:obs_p07}, but for the P07best$_{\rm MFL}$ (left) and P07ecc$_{\rm MFL}$ (right) model.}
\label{fig:obs_nodm_ml6}
\end{figure*}

So far we have assumed that the initial satellite's DM halo is
extended.  In principle, an alternative possibility is that, in the
absence of an extended DM halo, tidal forces are effective in
producing a flat and high LOS velocity-dispersion profile as observed
\citep{Aar83}. For instance, such a model appears to provide a good
match to the observed structure and kinematics of the Carina dSph
\citep{Mun08}.  To explore this idea for the case of Fornax, we ran
two additional simulations in which we make the extreme assumption
that ``mass follows light''.  In practice, in these simulations we do
not distinguish between stars and DM: the satellite is represented
initially as a one-component isotropic system of total mass
$\Mtot=1.5\times10^8\Msun$ with the same initial density profile as
the stellar component of simulations P07best and P07ecc. With this
choice the central LOS velocity dispersion is similar to the value
measured in Fornax. However, as expected, the initial
velocity-dispersion profile is much steeper than observed. We
considered both the P07best orbit (simulation P07best$_{\rm MFL}$) and
the more eccentric orbit P07ecc (simulation P07ecc$_{\rm MFL}$). The
resulting spatial distribution of particles after 2, 7 and 12 Gyr are
shown in Figs. \ref{fig:xyz_p07_nodm_ml6} and
\ref{fig:xyz_p07ecc_nodm_ml6}; the corresponding intrinsic density
profiles are shown in Fig. \ref{fig:rho_nodm_ml6}, and the
surface-density and LOS velocity-dispersion profiles in
Fig.~\ref{fig:obs_nodm_ml6}.

In simulation P07best$_{\rm MFL}$ a small fraction ($\sim$2\%) of the
initial mass appears to be decoupled from the main body of the galaxy,
thus forming a hint of tidal tails. At the end of the evolution both
the projected number-density and LOS velocity-dispersion profiles of
the stars are very similar to the initial ones.  Therefore, the LOS
velocity dispersion profile is still significantly decreasing with
radius, being just somewhat higher than the initial profile at $R>0.6$
deg, with velocity dispersion $\approx 6\kms$. This is at odds with
observations, in which the velocity dispersion profile is flat with
values $\approx 12$ km/s up to $R\simeq 0.7$ deg (see
Fig.~\ref{fig:obs_nodm_ml6}, left-hand panels). In simulation
P07ecc$_{\rm MFL}$ the system develops a clearer pair of tidal
tails. However, also in this case the satellite loses a small fraction
of the initial satellite mass budget ($\sim$3\%), so the final density
and velocity dispersion profiles are very similar to those of
simulation P07best$_{\rm MFL}$. High values of the final LOS velocity
dispersion are found in simulation P07ecc$_{\rm MFL}$ only at $R>1$
deg, far beyond the region covered by kinematic observations (see
Fig.~\ref{fig:obs_nodm_ml6}, right-hand panels). We note that in
  the orbits considered here the tidal tails developed by the
  satellite are never aligned along the LOS (a configuration that
  could enhance the LOS velocity dispersion at smaller projected
  radii; Klimentowski et al. 2007).

These results suggest that, even in the absence of an extended DM
halo, neither the large amplitude of Fornax's velocity-dispersion
profile, nor its approximately flat shape, are caused by lack of
dynamical equilibrium due to tidal disturbances from the MW.  In other
words, Fornax's kinematic samples are not expected to be significantly
contaminated by tidally stripped stars.

\subsection{A flattened model}
\label{sec:flattened}

In the simulations described in the previous sections we have assumed
that the satellite is initially spherical. In fact, we mentioned in
Section~\ref{sec:obs} that the observed stellar component of Fornax is
flattened, with observed ellipticity $\epsilon=0.3$. The simplest
interpretation of this deviation from circular symmetry is that
Fornax's stellar component is intrinsically axisymmetric. Given the
available observational information, the inclination of the system
with respect to the LOS is unknown and it is unclear how much rotation
or velocity anisotropy contribute to the observed flattening.  A more
realistic simulation of the evolution of Fornax in the MW should start
from a non-spherically symmetric satellite and match, in the final
snapshot, the observed ellipticity, together with the other
observables.  Clearly this is not an easy task, because intrinsic
shape, inclination, rotation and anisotropy add several degrees of
freedom. In the spirit of showing that the observed ellipticity can be
reproduced sufficiently well with a simple model and that the results
described above are not specific to spherically symmetric models, here
we present, as an example, the result of an additional simulation
named P07best$_{\rm flat}$, which follows the orbit P07best (see
Table~\ref{tab:sim}).

In simulation P07best$_{\rm flat}$ the satellite is initially
represented by an oblate (but non-rotating) stellar component and by
an almost-spherical DM halo. In practice, the initial conditions are
constructed as follows. We start by generating a two-component
spherical, isotropic system with stellar and DM density profiles given
by equations~(\ref{eq:serdeproj}) and (\ref{eq:rhodm}), respectively,
with the same parameters as in simulation P07best, but $\MDM=2.6\times
10^9\Msun$. Each particle is characterized by coordinates $x$, $y$,
$z$, $\vx$, $\vy$, $\vz$ centred in the dwarf galaxy's centre of
mass. The phase-space coordinates of the stellar\footnote{The
  phase-space coordinates of the DM particles are left unchanged.}
particles are modified as follows: $x$, $y$, $\vx$ and $\vy$ are
multiplied by a coefficient $a$, and $z$ and $\vz$ by a coefficient
$b<a$. In practice we use $a=q^{-1/3}$ and $b=q^{2/3}$, so the final
minor-to-major intrinsic axis ratio is $q=b/a$. Specifically, we adopt
$q=0.58$, so the (constant) intrinsic ellipticity is $\epsilon_{\rm i}
=1-q=0.42$. Such a system is not in equilibrium, but we verified
empirically that it is actually a quasi-equilibrium configuration. We
followed the evolution of such a system in isolation with an $N$-body
simulation, finding that an equilibrium configuration, still with
ellipticity $\epsilon_{\rm i}\simeq 0.41$, is reached after 1 Gyr of
evolution. The DM distribution slightly adjusts itself because of the
presence of the flattened stellar component, but we find that after 1
Gyr the halo is still very close to spherical symmetry.  We take as
initial ($t=0$) satellite in simulation P07best$_{\rm flat}$ this
$N$-body system as it is after 1 Gyr of evolution in isolation. The
initial phase-space coordinates of the centre of mass of the dwarf are
the same as in simulation P07best and the orientation of the system's
axes is empirically chosen so that at the end of the simulation the
system is seen almost edge-on when observed from the Sun. The
  iterative algorithm described in Section~\ref{sec:setup} has been
  used to find the optimal parameters at the beginning of the
  simulation, including the initial intrinsic ellipticity
  $\epsilon_{\rm i}$, which has been updated at each iteration
  according to the formula
  $$\epsilon_{\rm i}'= \epsilon_{\rm i}(\epsilon^{\rm
    obs}/\epsilon^{\rm sim})^{0.5},$$ where $\epsilon^{\rm obs}=0.3$
  and $\epsilon^{\rm sim}$ are the projected ellipticities calculated
  on observed data and in the last snapshot of the simulation (we took
  as initial guess $\epsilon_{\rm i}=0.3$). The initial orientation of
  the system is such that the equatorial plane is parallel to the
  LOS from the Galactic centre to the observed position of
  Fornax. During the simulation the vector normal to the satellite's
  equatorial plane changes its direction, oscillating by several
  degrees, but in the last snapshot the satellite is still seen from
  the Sun position as approximately edge-on.  The final properties of
simulation P07best$_{\rm flat}$ are reported in Table~\ref{tab:final}.

\begin{figure}
\includegraphics[width=8cm]{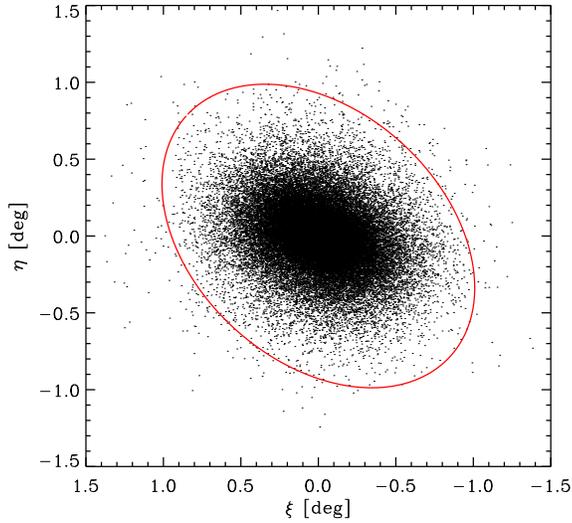}
\caption{Spatial distribution of stellar particles in simulation
  P07best$_{\rm flat}$ projected in the tangent plane as seen from the
  Sun. The ellipse has the ellipticity, position angle and nominal
  King tidal radius as in measured by B06 (see Table~\ref{tab:prop}).}
\label{fig:fov}
\end{figure}

\begin{figure}
\includegraphics[width=8cm]{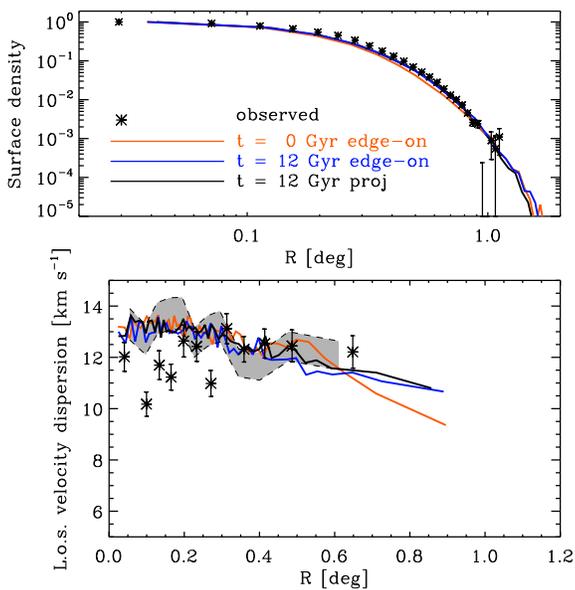}
\caption{As in Fig.~\ref{fig:obs_p07}, but for the P07best$_{\rm flat}$
  model. The black curves show the profiles as seen from the Sun.  The
  orange and blue lines show the profiles of the initial and final
  snapshots, as seen edge-on.}
\label{fig:obs_p07_fl}
\end{figure}

Figure~\ref{fig:fov} shows the distribution of stellar particles in
the tangent plane, centred at the coordinates of the object centre of
mass, as seen from the Sun.  The projected ellipticity of this
simulated object is $\epsilon\simeq 0.3$, in perfect agreement with
the value measured for Fornax. As it can be noted from
Fig.~\ref{fig:obs_p07_fl} both the resulting surface number density
and LOS velocity profile are also in this case in relatively good
agreement with Fornax's observed properties. The DM and stellar mass
loss is similar to the P07best spherical case: less than 3\% of stars
and $\sim$8\% of DM are lost within a radius of 1.6 kpc.  In
Fig.~\ref{fig:obs_p07_fl} we also overlay the profiles of the
simulated stellar component at the end of the evolution, as it would
appear if seen exactly edge-on: in this case the object's ellipticity
would be $\epsilon\simeq 0.42$, but the surface-density and LOS
velocity dispersion profiles would not be much different from those
found projecting along the actual LOS in the simulation.

\begin{table}
\centering
\caption{Final properties of the dwarf galaxy in the two-component
  simulations.  $M_{*,1.6}$: stellar mass within an angle-averaged
  radius $r=1.6\kpc$; $M_{\rm DM,1.6}$: DM mass within $r=1.6\kpc$;
  $f_{\rm *,1.6}$: final-to-initial stellar mass ratio within
  $r=1.6\kpc$; $f_{\rm DM,1.6}$: final-to-initial DM mass ratio within
  $r=1.6\kpc$.}
\label{tab:final}
\begin{tabular}{lcccc}
\hline
Name &  $M_{*,1.6}$ & $M_{\rm DM,1.6}$ & $f_{*,1.6}$  & $f_{\rm DM,1.6}$ \\
     &  $10^7\Msun$ & $10^8\Msun$  &  &  \\
\hline
P07best         &  4.62 & 2.64  & 0.991  & 0.985 \\
P07ecc          &  4.30 & 1.58  & 0.923  & 0.590 \\
P07ecc$_{\rm sh}$ & 4.70 & 2.95   & 0.975   & 0.787 \\
P07best$_{\rm flat}$ & 4.53 & 2.40 & 0.971 & 0.917 \\
\hline
\end{tabular}
\end{table}

\section{Discussion and Conclusions}
\label{sec:summary}

In this work we simulated the dynamical evolution over a 12 Gyr orbit
around the MW of a satellite, whose final structure and kinematics
resemble those of the Fornax dSph.  The results shown in
Section~\ref{sec:results} indicate that even assuming a relatively
eccentric orbit (still within 3$\sigma$ the best-fit systemic
velocities estimated by P07) the effects produced by the Galactic
tidal field are relatively small, and the final system is close to
equilibrium in its own gravitational potential. This suggests that the
assumption of dynamical equilibrium can be safely used to estimate the
mass (and DM content) of Fornax and that the existing kinematic
samples are not expected to be contaminated by tidally stripped stars.

Our result is robust in the sense that our models were built so to
maximize the possible tidal effects, because we followed Fornax orbit
for as long as 12 Gyr in a ``fully grown'' MW, and we considered
  the orbits with the smallest pericentre still consistent with direct
  measurements of Fornax proper motion. According to hierarchical
models of galaxy formation, a system like the MW is expected to have
substantially increased its DM mass over 12 Gyr and, of course, also
the MW stellar components have not been in place as they are now for
such a long time span.  Furthermore, Fornax may have become a MW
satellite more recently than 12 Gyr ago.  For example, \citet{Roc12},
based on the results of the Via Lactea~II simulation, propose an
infall time for Fornax of 5-9 Gyr ago. Another indication that the
orbital evolution of Fornax in the MW was likely shorter than 12 Gyr
can be found in its very extended star formation history, with star
formation having occurred until $\sim$50-100 Myr ago.  It is still
debated on what timescales ram-pressure stripping can remove the
gaseous component of the MW dSphs progenitors; nonetheless, the
observational finding that Local Group gas-rich star-forming dwarfs
are mostly located beyond the estimated MW or M31 virial radius
(e.g. van den Bergh 2000, Irwin et al. 2007) suggests that objects
with recent star formation activity must have spent a considerable
fraction of their lifetime far from the large Local Group spirals.
Overall, we expect that in a more realistic set-up the effect of tides
would be milder than what examined here.

As we describe the MW as a fixed gravitational potential, we neglect
the effect of dynamical friction on the satellite's orbit. As Fornax
orbits in relatively low-density regions of the MW, dynamical friction
is not expected to be very important. In any case, the effect of
dynamical friction is to bring the satellite closer to the host's
centre: if dynamical friction were accounted for, the backward in time
reconstructed orbit of Fornax would be more extended than the one
considered here. Also in this sense, our simulations with fixed MW
potential tend to slightly overestimate the tidal effects.
Moreover, a more external orbit and consequently weaker tidal effects would be obtained by assuming a less
  massive model for the MW, as recently suggested by Gibbons,
  Belokurov \& Evans (2014) on the basis of the shape and extent of
  the Sagittarius stream.

Based on the results of the present work, we are reassured on the fact
that dynamical mass estimates of Fornax derived from equilibrium
dynamical models are reliable.  Still, as well known, dynamical models
suffer from the mass-anisotropy degeneracy, so that different
combinations of DM density and stellar-anisotropy profiles can
reproduce equally well the observations. Moreover, given Fornax's
observed ellipticity, additional degrees of freedom derive from the
uncertainties on the intrinsic shape and rotational support of this
dSph (see \citealt{Hay12} and \citealt{Jar12}, who studied
axisymmetric equilibrium models of Fornax).  However, while the inner
DM distribution is hard to determine, the mass within 1.6 kpc
(approximately the outermost point of the observed LOS velocity
dispersion profile of Fornax) is believed to be more robustly
constrained (BH13).  Therefore, in the limits of the considered
approximations (initial spherical symmetry and isotropic velocity
distribution, with the exception of simulation P07best$_{\rm flat}$),
our simulations can be also used to estimate Fornax's DM mass over the
scales probed by the existing kinematic data-sets. The simulations
that successfully reproduce Fornax's observables have a
dark-to-luminous mass ratio within 1.6 kpc in the range $5-6$ (see
Table~\ref{tab:final}).  Measuring the same ratio within 3 kpc yields
values in the range $16-18$. We confirm that Fornax is amongst the
least DM dominated dSphs, and we suggest that baryons and DM may
contribute similarly to the mass budget in Fornax's central
regions. In the near future we plan to extend the present analysis to
other dSphs for which the orbital parameters are sufficiently well
constrained.

\section*{Acknowledgments}

The authors thank M. Breddels for providing Fornax's LOS velocity
dispersion profile measured in elliptical annuli and the referee, M.
Wilkinson, for useful comments.  GB thanks the INAF - Bologna for the
hospitality during part of this work and gratefully acknowledges
support through a Marie-Curie action Intra European Fellowship, funded
by the European Union Seventh Framework Program (FP7/2007-2013) under
Grant agreement number PIEF-GA-2010-274151, as well as the financial
support by the Spanish Ministry of Economy and Competitiveness
(MINECO) under the Ramon y Cajal Programme (RYC-2012-11537). AS
acknowledges financial support from PRIN INAF 2011 ``Multiple
populations in globular clusters: their role in the Galaxy assembly''
(PI E. Carretta). CN thanks the Instituto de Astrofisica de Canarias
for the hospitality and acknowledges financial support from PRIN MIUR
2010-2011, project ``The Chemical and Dynamical Evolution of the Milky
Way and Local Group Galaxies'', prot. 2010LY5N2T.

\clearpage

\bsp

%


\label{lastpage}

\end{document}